\documentclass{aa}  

\usepackage{graphicx}
\usepackage{txfonts}
\usepackage{natbib}
\usepackage{longtable}
\usepackage{subfigure}
\usepackage{threeparttable}
\begin{document}

\authorrunning{A. Bonafede et al.}  \titlerunning{Fractional polarization and magnetic fields in the ICM} \title{Fractional polarization as a
  probe of magnetic fields in the intra-cluster medium}
\subtitle{ }
\author{A. Bonafede \inst{1,2,3} \and F. Govoni\inst{4} \and
  L. Feretti \inst{3} \and M. Murgia \inst{4} \and G. Giovannini
  \inst{2,3} \and M. Br\"uggen \inst{1} }

   \offprints{a.bonafede@jacobs-university.de} \institute{Jacobs
     University Bremen, Campus Ring 1, D-28759 Bremen, Germany
     \\ \and Dip. di Astronomia, Univ. Bologna, via Ranzani 1, I-40120
     Bologna, Italy\\ \and INAF- Istituto di Radioastronomia, via
     Gobetti 101, I-40129, Italy \\ \and INAF- Osservatorio
     Astronomico di Cagliari,Loc. Poggio dei Pini, Strada 54, I-09012,
     Capoterra (Ca) Italy \\}

   \date{Received ; accepted }
% 5 {} token are mandatory
  \abstract 
% context heading (optional) % {} leave it empty if necessary 
{It is
  now established that magnetic fields are present in the
  intra-cluster medium (ICM) of galaxy clusters, as revealed by
  observations of radio halos and radio relics and from the study of
  the Faraday Rotation Measures of sources located either behind or
  within clusters. Deep radio polarization observations of clusters
  have been performed in the last years, and the properties of the ICM
  magnetic field have been constrained in a small number of
  well-studied objects.}
% aims heading (mandatory) 
{ The aim of this work is to investigate the average properties of the
  ICM magnetic fields, and to search for possible correlations with
  the ICM thermal properties and cluster radio emission. }
% methods heading (mandatory) 
{We have selected a sample of 39 massive galaxy clusters from the
  HIghest X-ray FLUx Galaxy Cluster Sample, and used Northern VLA Sky
  Survey data to analyze the fractional polarization of radio sources
  out to 10 core radii from the cluster centers. Following Murgia et
  al (2004), we have investigated how different magnetic field
  strengths affect the observed polarized emission of sources lying
  at different projected distances from the cluster center. In
  addition, statistical tests are performed to investigate the
  fractional polarization trends in clusters with different thermal
  and non-thermal properties.}
% results heading (mandatory) 
{We find a trend of the fractional
polarization with the cluster impact parameter, with fractional
polarization increasing at the cluster periphery and decreasing toward
the cluster center. Such trend can be reproduced by a magnetic field
model with central value of few $\mu$G. The logrank statistical test
indicates that there are no differences in the depolarization trend
observed in cluster with and without radio halo, while the same test
indicates significant differences when the depolarization trend of
sources in clusters with and without cool core are compared. The
comparison between clusters with high and low temperatures does not
yields significant differences. Although the
role of the gas density should be better accounted for, these results
give important indications for models that require a role of the ICM
magnetic field to explain the presence of cool core and radio halos in
galaxy clusters.}  {} \keywords{Cluster of galaxies -- Magnetic field
    -- Depolarization --- non-thermal }
   
\maketitle
%
%________________________________________________________________

\section{Introduction}
Increasing attention has been devoted in the last decade to the 
presence, strength and structure of magnetic fields in galaxy clusters
(\citealt{2008SSRv..134...93F}, \citealt{2002ARA&A..40..319C},
\citealt{2004IJMPD..13.1549G}, \citealt{2004JKAS...37..337C} ). In
order to understand the physical conditions of the intra cluster
medium (ICM) it is important to understand the properties of cluster
magnetic fields and their interplay with the other constituents of the
ICM. Our current knowledge on cluster magnetic fields comes mainly
from radio observations. In some galaxy clusters magnetic fields are
revealed by synchrotron emission not obviously associated with any
particular galaxy: the so-called radio halos
(e.g. \citealt{2009A&A...507.1257G}). Radio halos are wide synchrotron
radio sources, characterized by steep spectra\footnote{$S(\nu)\propto
  \nu^{-\alpha}$, with $\alpha=$ the spectral index} ($\alpha >$1) and
low surface brightness ($\sim$ 1 $\mu$Jy/arcsec$^2$ at 1.4 GHz). They
are located at the center of galaxy clusters, and are usually found to
be unpolarized, the only two exceptions being the cluster Abell 2255
(\citealt{2005A&A...430L...5G}, see also
\citealt{2010arXiv1008.3530P}) and MACS J0717+3745
\citep{2009A&A...503..707B}.  Synchrotron diffuse emission is also
observed at the periphery of some galaxy cluster (the so called radio
relics, see e.g. \citealt{2009A&A...494..429B},
\citealt{2010Sci...330..347V}) About 30 radio halos are known so far,
and all of them are found in clusters with clear signature of on-going
or recent merger activity (e.g. \citealt{2001ApJ...553L..15B},
\citealt{2001A&A...369..441G}, \citealt{2010ApJ...721L..82C}). Shocks
and turbulence associated with merger events are expected to inject a
considerable amount of energy in the ICM, that could compress and
amplify the magnetic field and accelerate relativistic particles,
giving thus rise to the observed radio emission.  We refer to
\citet{2008SSRv..134...93F} and \citet{2008SSRv..134..311D} for recent
reviews of the subject.\\ Estimates of the magnetic field strength
have been obtained in clusters where radio halos are observed, under
the minimum energy hypothesis (which is very close to equipartition
conditions) or by studying the Inverse Compton (IC) Hard X-ray
emission (e.g. \citealt{2004Ap&SS.294...37F},
\citealt{2009ApJ...696.1700W}, \citealt{2009ApJ...690..367A},
\citealt{2010A&A...514A..76M}). Both of these methods require however
several assumptions on the emitting particle distribution and
energetic spectrum. In particular the assumptions required by the
equipartition approach, in the context of radio halos, strongly affect
the resulting estimates. Magnetic field values derived from
equipartition and IC indicate that magnetic of $\sim\mu$G level are
spread over the cluster volume.  Recently, a more sophisticated
approach has been applied by \citet{2010A&A...514A..71V}. The authors
investigated the magnetic field power spectrum in the cluster A665 by
analyzing the radio halo brightness fluctuations in conjunction with
the fractional polarization trend of radio sources at different impact
parameters. \\ Another possibility to investigate the ICM magnetic
field properties comes from the study of the Faraday rotation of
sources located both behind and within galaxy clusters.
Synchrotron radiation from radio galaxies which crosses a magneto-ionic
medium is subject to Faraday Rotation. The direction
of the polarization plane, $\Psi_{int}$, is rotated by a quantity that
in the case of a purely external Faraday screen is proportional to the
square of the wavelength:
\begin{eqnarray}
&& \Psi_{obs}(\lambda)=\Psi_{int}+RM\lambda^2 \\
&&  RM \propto \int_0^L{B_{//}n_g dl},
\label{eq:psiobs}
\end{eqnarray}
here $B_{//}$ is the magnetic field component along the line of sight,
$n_g$ is the ICM gas density and $L$ is the distance along the line of
sight. With the help of X-ray observations, providing information
about the thermal gas distribution, $RM$ studies give an additional
set of information about the magnetic field in the ICM.  Recent works
have investigated several aspects of the magnetic field morphology,
such as its power spectrum (\citealt{2004A&A...424..429M},
\citealt{2006A&A...460..425G}, \citealt{2010A&A...513A..30B},
\citealt{2005A&A...434...67V}, \citealt{2008MNRAS.391..521L}) and its
central strength and radial decline \citep{2010A&A...513A..30B}. These
studies, however, require deep and multi-frequency observations of
several sources located at different projected distances from the
cluster center, and have thus been performed so far on a small number
of clusters. Because of the sensitivity limits of radio telescopes,
studies of a large number of galaxy clusters with many $RM$ probes per
cluster are still unfeasible with the current instruments. Hence in
order to obtain general information on magnetic fields in galaxy
clusters without focusing on single objects, another strategy is
required.\\ When synchrotron emission arising from a cluster or
background source crosses the ICM, regions with similar $\Psi_{int}$,
going through different paths, will be subject to differential Faraday
Rotation. If the magnetic field in the foreground screen is tangled on
scales much smaller than the observing beam, radiation with similar
$\Psi$ but opposite orientation will averaged out, and the observed
degree of polarization will be reduced (beam depolarization). In the
central region of a cluster, the magnetic field is expected to be
higher, according to both theoretical studies (see
\citealt{2008SSRv..134..311D} for a review) and Faraday RM observation
(\citealt{2004JKAS...37..337C}, \citealt{2006A&A...460..425G},
\citealt{2008A&A...483..699G}, \citealt{2010A&A...513A..30B}). Higher
value of $B$ and $n_g$ result in higher $RM$, according to
Eq. \ref{eq:psiobs}. The higher the Faraday RM is, the stronger the
beam-depolarization. It is then expected that sources located in
projection close to the cluster center will show a lower fractional
polarization compared to more distant
ones.\\ \citet{2004A&A...424..429M} used the FARADAY code to simulate
this effect, proving that a large sample of radio sources could give
additional statistical constraints on the intra-cluster magnetic field
properties. The advantage of this approach is that there is no need
for multi-frequency observations. This effect can be investigated with
radio surveys at a single frequency, performed in full polarization
mode. Such studies do not provide detailed information on magnetic
fields in specific clusters, but allow us to understand the average
properties of magnetic fields in the ICM. \\ In this paper, we present
a statistical study of magnetic fields in galaxy clusters. Starting
from a sample of clusters selected from the HIghest X-ray FLUx Galaxy
Cluster Sample (HIFLUGCS, \citealt{2002ApJ...567..716R}), we use the
NRAO Northern VLA Sky Survey (NVSS, \citealt{1998AJ....115.1693C}) to
study the polarization properties of sources located at different
impact parameters with respect to the cluster center.\\ The paper is
organized as follows: In Sec. \ref{sec:HIFLUGCS} we present the
cluster sample, in Sec. \ref{sec:radio} radio data are analyzed; The
polarization properties of the radio sources are studied in
Sec. \ref{sec:DP}, and used to derive the average magnetic field
properties (Sec. \ref{sec:Bprop}) with the help of numerical
simulations. In Secs.  \ref{sec:halo}, \ref{sec:coolcore} and
\ref{sec:BT} the difference in magnetic fields in clusters with and
without radio halos and with high and low temperature are
investigated. Finally, conclusions are presented in
Sec. \ref{sec:concl}. In this work we assume a $\Lambda$CDM
cosmological model, characterized by $\Omega_{\Lambda}=0.7$,
$\Omega_m=0.3$ and $H_0=$70 km/s/Mpc.

\section{Selection of the clusters}
\label{sec:HIFLUGCS}
From the HIghest X-ray FLUx Galaxy Cluster Sample (HIFLUGCS) by
\citet{2002ApJ...567..716R}, we selected all objects with $L_x[0.1-2.4
  keV]\geq 1.5 \times 10^{44}$erg/s. The limit in X-ray luminosity is
aimed at selecting massive galaxy clusters, and at the same time
having enough clusters to build up a statistical sample. In massive
galaxy clusters the magnetic field is expected to be higher, so that
the depolarization effect should be more prominent. There are 39
clusters in the HIFLUGCS sample that satisfy this criterion. Using the
$L_X-M_{200}$ relation obtained by \citet{2008MNRAS.387L..28R} this
limit corresponds to $M_{200} \geq 3\times 10^{14}M_{\odot}/h_{70}$.
The HIFLUGCS sample is a X-ray selected and X-ray flux-limited galaxy
cluster sample from the ROSAT All-Sky-Survey catalogue. The large FOV
of the ROSAT PSPC covers most of the clusters out to $r_{500}$, which
is the radius at which the mean density of the cluster is 500 times
that of a critical density. The selection criterion that we used,
based on the cluster X-ray luminosity only, guarantees that clusters
in different dynamical states are included in our sample. \\ The
HIFLUGCS sample has been used to perform several studies regarding the
scaling relations of galaxy clusters ({\it e. g.} 
\citealt{2007A&A...466..805C}). Their surface brightness profiles have
been derived from pointed ROSAT PSPC (32 clusters) and RASS (7
clusters) observations \citep{2002ApJ...567..716R} by using the
standard $\beta-$model \citep{1976A&A....49..137C}. From X-ray
observations the cluster mean temperature T and the gas density
distribution have been derived. The gas density distribution is
described by the following equation:
\begin{equation}
\label{eq:Sxbeta}
n_g(r)=n_0\left( 1+\frac{r^2}{r_c^2} \right)^{-\frac{3}{2}\beta},
\end{equation}
here $r$ is the distance from the cluster center and $r_c$ is the
cluster core radius.\\ The basic properties of the clusters in the
sample, taken from \citet{2002ApJ...567..716R} and rescaled to the
cosmological model adopted in this work, are given in Tables
\ref{tab:sample1} and \ref{tab:sample2} . In the same Table, we list
the dynamical state of the clusters as found in the literature,
indicating clusters with merging signature (M), and with mild or
strong cool core (CC).

\begin{table*} [h]
\caption{Cluster sample selected from the HIFLUGCS catalogue}        
\label{tab:sample1}      
\centering          
%\begin{threeparttable}
\begin{tabular}{|c c c c c |}     % 8 columns 
\hline
\hline       
Cluster & RA & DEC & z & Angular to linear conversion scale \\
name &   (J2000)& (J2000)&& $''$/kpc\\
\hline                  
&&&&\\
     2A0335 & 03h38m35.3 & +09d57m55s &  0.0349 &  0.6950 \\
     A0085 & 00h41m37.8 & -09d20m33s &  0.0556 &  1.0800 \\
     A0119 & 00h56m21.4 & -01d15m47s &  0.0440 &  0.8660 \\
     A0133 & 01h02m39.0 & -21d57m15s &  0.0569 &  1.1030 \\
     A0399 & 02h57m56.4 & +13d00m59s &  0.0715 &  1.3630 \\
     A0401 & 02h58m56.9 & +13d34m56s &  0.0748 &  1.4240 \\
     A0478 & 04h13m20.7 & +10d28m35s &  0.0900 &  1.6790 \\
     A0496 & 03h51m58.6 & -22d10m05s &  0.0328 &  0.6550 \\
     A0754 & 09h08m50.1 & -09d38m12s &  0.0528 &  1.0290 \\
     A1644 & 12h57m14.8 & -17d21m13s &  0.0474 &  0.9300 \\
     A1650 & 12h58m46.2 & -01d45m11s &  0.0845 &  1.5870 \\
     A1651 & 12h59m22.9 & -04d11m10s &  0.0860 &  1.6120 \\
     A1736 & 13h26m52.1 & -27d06m33s &  0.0461 &  0.9050 \\
     A1795 & 13h49m00.5 & +26d35m07s &  0.0616 &  1.1880 \\
     A2029 & 15h10m56.0 & +05d44m41s &  0.0767 &  1.4530 \\
     A2065 & 15h22m42.6 & +27d43m21s &  0.0721 &  1.3730 \\
     A2142 & 15h58m16.1 & +27d13m29s &  0.0899 &  1.6770 \\
     A2147 & 16h02m17.2 & +15d53m43s &  0.0351 &  0.6980 \\
     A2163 & 16h15m34.1 & -06d07m26s &  0.2010 &  3.3130 \\
     A2199 & 16h28m38.5 & +39d33m06s &  0.0302 &  0.6050 \\
     A2204 & 16h32m45.7 & +05d34m43s &  0.1523 &  2.6480 \\
     A2244 & 17h02m44.0 & +34d02m48s &  0.0970 &  1.7950 \\
     A2255 & 17h12m31.0 & +64d05m33s &  0.0800 &  1.5100 \\
     A2256 & 17h03m43.5 & +78d43m03s &  0.0601 &  1.1610 \\
     A2597 & 23h25m18.0 & -12d06m30s &  0.0852 &  1.5980 \\
     A3558 & 13h27m54.8 & -31d29m32s &  0.0480 &  0.9410 \\
     A3562 & 13h33m31.8 & -31d40m23s &  0.0499 &  0.9760 \\
     A3571 & 13h47m28.9 & -32d51m57s &  0.0397 &  0.7860 \\
     A4059 & 23h56m40.7 & -34d40m18s &  0.0460 &  0.9040 \\
      COMA & 12h59m48.7 & +27d58m50s &  0.0232 &  0.4680 \\
   HYDRA-A & 09h18m30.3 & -12d15m40s &  0.0538 &  1.0470 \\
     MKW3S & 15h21m50.0 & +07d42m32s &  0.0450 &  0.8850 \\
  ZwCl1215 & 12h17m41.4 & +03d39m32s &  0.0750 &  1.4240 \\
\hline

\multicolumn{5}{l}{\scriptsize Col. 1: Cluster name; Col. 2, Col. 3:
  Cluster center (RA, DEC); Col. 4 : redshift; Col. 5: angular scale;}\\
\end{tabular}
\end{table*}

\begin{table*} [h]
\caption{Cluster sample: Thermal properties}        
\label{tab:sample2}      
\centering          
\begin{threeparttable}
\begin{tabular}{|c c c c c c c c |}     % 8 columns 
\hline
\hline       
Cluster & L$_{x}$ [0.1 - 2.4 keV] & r$_c$ & n$_0$ & $\beta$ & T & Radio & State \\
name    &  10$^{44}$ erg/s        &  kpc  &  10$^{-2}$cm$^{-3}$& & keV&&   \\
\hline                  
&&&&&&&\\
     2A0335 &  2.53 &    24$^{+  0}_{-  0}$ & 5.29$^{+0.06}_{-0.06}$ & 0.575$^{0.004}_{0.003}$& 3.01$^{+0.07}_{-0.07}$ & &CC\tnote{a} \\
     A0085 &  5.28 &    60$^{+  2}_{-  2}$ & 2.43$^{+0.09}_{-0.09}$ & 0.532$^{0.004}_{0.004}$ & 6.10$^{+0.20}_{-0.20}$ & &CC\tnote{a}\\
     A0119 &  1.79 &   365$^{+ 19}_{- 18}$ & 0.14$^{+0.01}_{-0.01}$ & 0.675$^{0.026}_{0.023}$ & 5.80$^{+0.60}_{-0.60}$ & &M\tnote{b} \\
     A0133 &  1.59 &    33$^{+  0}_{-  0}$ & 2.65$^{+0.08}_{-0.08}$ & 0.530$^{0.004}_{0.004}$ & 3.80$^{+2.00}_{-0.90}$ & &CC\tnote{a}\\
     A0399 &  3.87 &   332$^{+ 96}_{- 73}$ & 0.21$^{+0.04}_{-0.03}$ & 0.713$^{0.0137}_{0.095}$& 7.40 $^{+0.70}_{-0.70}$&H\tnote{h1} &M \tnote{p} \\
     A0401 &  6.90 &   182$^{+  8}_{-  7}$ & 0.56$^{+0.05}_{-0.04}$ & 0.613$^{0.010}_{0.010}$ & 8.30$^{+0.50}_{-0.50}$ &H\tnote{h2} &M \tnote{P}\\
     A0478 &  9.86 &    73$^{+  1}_{-  1}$ & 3.26$^{+0.14}_{-0.13}$ & 0.613$^{0.004}_{0.004}$ & 7.10$^{+0.40}_{-0.40}$ & &CC\tnote{a}\\
     A0496 &  2.03 &    21$^{+  0}_{-  0}$ & 3.94$^{+0.28}_{-0.25}$ & 0.484$^{0.003}_{0.003}$ & 4.13$^{+0.08}_{-0.08}$ & &CC\tnote{a}\\
     A0754 &  2.15 &   175$^{+ 12}_{- 11}$ & 0.42$^{+0.02}_{-0.02}$ & 0.698$^{0.027}_{0.024}$ & 9.00$^{+0.05}_{-0.05}$ &H\tnote{h3}& M\tnote{o}\\
     A1644 &  2.07 &   218$^{+ 93}_{- 67}$ & 0.27$^{+0.08}_{-0.06}$ & 0.579$^{0.111}_{0.074}$ & 4.70$^{+0.90}_{-0.70}$ & & M\tnote{c}\\
     A1650 &  4.05 &   209$^{+ 77}_{- 52}$ & 0.40$^{+0.08}_{-0.06}$ & 0.704$^{0.131}_{0.081}$ & 5.60$^{+0.60}_{-0.60}$ & & D \tnote{q}\\
     A1651 &  4.44 &   134$^{+  6}_{-  6}$ & 1.01$^{+0.06}_{-0.06}$ & 0.643$^{0.014}_{0.013}$ & 6.30$^{+0.50}_{-0.50}$ & & CC\tnote{a}\\
     A1736 &  1.72 &   273$^{+129}_{- 94}$ & 0.12$^{+0.04}_{-0.02}$ & 0.542$^{0.147}_{0.092}$ & 3.50$^{+0.40}_{-0.40}$ & &M\tnote{d}\\
     A1795 &  5.49 &    57$^{+  0}_{-  0}$ & 2.71$^{+0.05}_{-0.05}$ & 0.596$^{0.003}_{0.002}$ & 6.00$^{+0.30}_{-0.30}$ & &CC\tnote{a}\\
     A2029 &  9.53 &    61$^{+  1}_{-  1}$ & 3.62$^{+0.14}_{-0.14}$ & 0.582$^{0.004}_{0.004}$ & 8.70$^{+0.30}_{-0.30}$ & &CC\tnote{a}\\
     A2065 &  3.05 &   509$^{+266}_{-137}$ & 0.19$^{+0.07}_{-0.04}$ & 1.162$^{0.737}_{0.282}$ & 5.40$^{+0.30}_{-0.30}$ & &M\tnote{e}\\
     A2142 & 11.89 &   114$^{+  3}_{-  3}$ & 1.48$^{+0.06}_{-0.06}$ & 0.591$^{0.006}_{0.006}$ & 8.80$^{+0.60}_{-0.60}$ & &CC\tnote{a}\\ 
     A2147 &  1.54 &   172$^{+ 74}_{- 46}$ & 0.16$^{+0.04}_{-0.03}$ & 0.444$^{0.071}_{0.046}$ & 4.91$^{+0.28}_{-0.28}$ & &M\tnote{f}\\
     A2163 & 20.87 &   405$^{+ 23}_{- 22}$ & 0.44$^{+0.02}_{-0.02}$ & 0.796$^{0.030}_{0.028}$ & 13.29$^{+0.64}_{-0.64}$& H\tnote{h5} & M\tnote{h5}\\
     A2199 &  2.19 &   100$^{+  7}_{-  6}$ & 0.81$^{+0.03}_{-0.03}$ & 0.655$^{0.019}_{0.021}$ & 4.10$^{+0.08}_{-0.08}$ & &CC\tnote{a}\\
     A2204 & 15.84 &    51$^{+  2}_{-  1}$ & 4.40$^{+0.10}_{-0.10}$ & 0.597$^{0.008}_{0.007}$ & 7.21$^{+0.25}_{-0.25}$ & &CC\tnote{a}\\
     A2244 &  4.75 &    93$^{+  7}_{-  7}$ & 1.09$^{+0.05}_{-0.05}$ & 0.607$^{0.016}_{0.015}$ & 7.10$^{+5.00}_{-2.20}$ & &CC\tnote{a}\\
     A2255 &  3.04 &   440$^{+ 26}_{- 23}$ & 0.17$^{+0.02}_{-0.02}$ & 0.797$^{0.033}_{0.030}$ & 6.87$^{+0.20}_{-0.20}$ & H\tnote{h4}& M\tnote{i} \\
     A2256 &  5.05 &   432$^{+ 28}_{- 26}$ & 0.25$^{+0.01}_{-0.01}$ & 0.914$^{0.054}_{0.047}$ & 7.50$^{+0.40}_{-0.40}$ & H\tnote{h6}&M\tnote{l}\\
     A2597 &  3.82 &    42$^{+  1}_{-  1}$ & 3.34$^{+0.07}_{-0.06}$ & 0.633$^{0.008}_{0.008}$ & 3.60$^{+0.20}_{-0.20}$ & &CC\tnote{a}\\
     A3558 &  3.54 &   163$^{+  3}_{-  3}$ & 0.44$^{+0.01}_{-0.01}$ & 0.580$^{0.006}_{0.005}$ & 5.50$^{+0.30}_{-0.30}$ & &M\tnote{g}\\
     A3562 &  1.67 &    71$^{+  3}_{-  3}$ & 0.55$^{+0.02}_{-0.02}$ & 0.472$^{0.006}_{0.006}$ & 5.16$^{+0.16}_{-0.16}$ & H\tnote{h7}&M\tnote{m}\\
     A3571 &  4.32 &   131$^{+  4}_{-  4}$ & 1.05$^{+0.09}_{-0.08}$ & 0.613$^{0.010}_{0.010}$ & 6.90$^{+0.30}_{-0.30}$ & &CC\tnote{a}\\
     A4059 &  1.54 &    65$^{+  3}_{-  3}$ & 1.13$^{+0.08}_{-0.08}$ & 0.582$^{0.001}_{0.001}$ & 4.10$^{+0.30}_{-0.30}$ & &CC\tnote{a}\\
      COMA &  4.13 &   247$^{+ 15}_{- 14}$ & 0.29$^{+0.06}_{-0.06}$ & 0.654$^{0.019}_{0.021}$ & 8.38$^{+0.34}_{-0.34}$ &H\tnote{h8} &M\tnote{n}\\
   HYDRA-A &  3.19 &    36$^{+  0}_{-  0}$ & 3.40$^{+0.35}_{-0.32}$ & 0.573$^{0.003}_{0.003}$ & 3.80$^{+0.20}_{-0.20}$ & &CC\tnote{a}\\
     MKW3S &  1.53 &    48$^{+  1}_{-  1}$ & 1.81$^{+0.24}_{-0.24}$ & 0.581$^{0.008}_{0.007}$ & 3.50$^{+0.20}_{-0.20}$ & &CC\tnote{a}\\
  ZwCl1215 &  2.88 &   319$^{+ 20}_{- 18}$ & 0.25$^{+0.01}_{-0.01}$ & 0.819$^{0.038}_{0.034}$ & 5.58$^{+0.89}_{-0.78}$ & & M\tnote{h}\\
&&&&&&&\\
\hline

\multicolumn{8}{l}{\scriptsize Col. 1: Cluster name; Col. 2: X-ray
  luminosity in the 0.1- 2.4 keV band; Col. 3: Cluster core radius;
  Col. 4: central density;}\\ \multicolumn{8}{l}{\scriptsize Col. 5:
  $\beta$ parameter; Col. 6: cluster mean temperature; Col. 7: Radio
  emission. H=giant radio halo detected; Col. 8: Dynamical
  State.}\\ \multicolumn{8}{l}{\scriptsize CC=cool-core cluster,
  M=merging system, D= debated, see explanation
  below.}\\ \multicolumn{8}{l}{\scriptsize If not specified, data are
  from Chen et al. 2006. Quantities have been corrected for the
  cosmological model assumed in this paper. }
\end{tabular}
\begin{tablenotes}
       \item[a] \citet{2007A&A...466..805C}, \item[b]
         \citet{2003AAS...203.4709W}, \item[c]
         \citet{2004ApJ...608..179R}, \item[d]
         \citet{2009POBeo..86..347B}, \item[e]
         \citet{2008A&A...479..307B}, \item[f]
         \citet{2002BaltA..11..269K} \& \citet{1997AJ....113.1939D}, \item[g]
         \citet{2007A&A...463..839R}, \item[h]
         \citet{2006AJ....132.1275R},
         \item[i]\citet{2003AJ....125.2427M},
           \item[l]\citet{2002ApJ...565..867S},
        \item[m]\citet{2004ApJ...611..811F},
          \item[n]\citet{2003A&A...400..811N},
          \item[o]\citet{2003ApJ...586L..19M}
            \item[p] \citet{2008A&A...479..307B},
              \item[q] no cool core according to Chen et al. (2006), but
                the cluster is found to be in a relaxed state with a
                moderate cool core, by the analysis of \citet{2003PASJ...55.1105T}
         \item[h1] \citet{2010A&A...509A..86M},
         \item[h2] \citet{2003A&A...400..465B},
         \item[h4] \citet{1980A&AS...39..215H},
         \item[h3] \citet{2001ApJ...559..785K},
         \item[h5] \citet{1994AAS...185.5307H}, \citet{2001A&A...373..106F}
         \item[h6] \citet{1976A&A....52..107B}
         \item[h7] \citet{2000MNRAS.314..594V}
          \item[h8] \citet{1980AJ.....85..183H}
     \end{tablenotes}
\end{threeparttable}
\end{table*}

\section{Radio Data}
\label{sec:radio}
Radio images of the selected clusters were retrieved from the Northern
VLA Sky Survey \citep{1998AJ....115.1693C}. The NVSS was performed at
1.4 GHz, with a 100 MHz bandwidth, and covers the sky north of $\delta
=$ -40 deg. The survey was performed in full polarization mode, so
that the total intensity radio flux (Stokes I) and Stokes Q and U
images are provided. The images all have a resolution of $\theta =
45''$ at the FWHM and nearly uniform sensitivity. Their noise rms
brightness fluctuations are $\sigma_I\sim $ 0.45 mJy/beam (Stokes I)
and $\sigma_{U,Q} \sim 0.30$ mJy/beam in Stokes Q and U. Radio images
for each cluster are centered on the cluster X-ray peak and have sizes
of 20 $r_c\times$ 20 $r_c$. \\The clusters S1101, A3391, A3667, A3266,
A3158 and A3112 lying at $\delta <$40 deg are not in the NVSS, so that
the radio sample consists finally of 33 clusters. Coma is the nearest
cluster of the sample, with $z=0.0232$, while Abell 2163 is the most
distant one, with $z=0.2010$. Given the resolution of the NVSS this
translates into a linear resolution ranging from 21 kpc to 150 kpc. This
guarantees that effects due to the beam depolarization should be
visible at the NVSS resolution, provided that the magnetic field is
tangled on scales smaller than these.
Except of Abell 2163 and Abell 2204, the clusters lie all at
z$<$0.1, their mean redshift is 0.066 with a rms dispersion of 0.032.
\\ From Stokes U and Q images the polarization intensity and
polarization angle images were produced:
\begin{eqnarray}
&& P=\sqrt{U^2+Q^2} \nonumber\\
&& \Psi = \frac{1}{2} arctan{\frac{U}{Q}}, 
\label{eq:UQ}
\end{eqnarray}
where $P$ was corrected for the positive bias.\\ Although the NVSS radio
images have a nearly uniform sensitivity, local regions of poorer
sensitivity may be present for instance because of dynamic range
limitations. In order to avoid including in our source sample peaks of
the noise fluctuations, only sources brighter than 5$\sigma_I$ in
total intensity images were considered. We blanked the total intensity
images at 5$\sigma_I$, and used these blanked images as a mask to
blank the polarization intensity images. We considered only sources
that are wider than one beam area. This prevents us
from including only the brightest part of low brightness sources,
which could bias the sample toward higher $F_P$ values. We then
computed the mean polarized intensity flux and the mean total
intensity flux for each source, and derived the fractional
polarization as $F_P= P/ I$.

\begin{figure}[ht]
\centering
\includegraphics[width=\columnwidth]{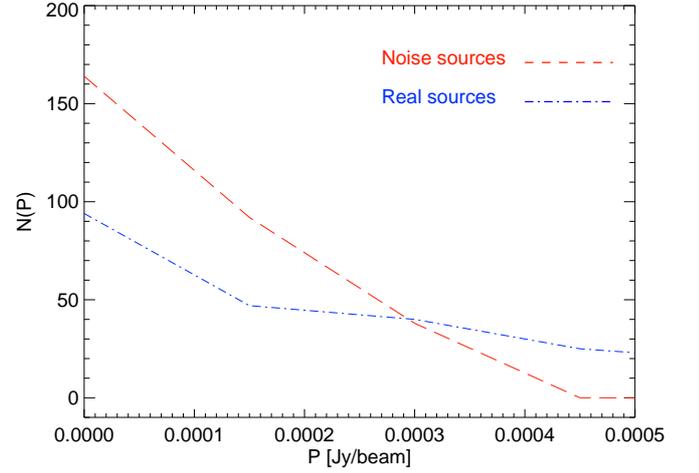}
\caption{Number of sources and noise as a function of the observed
  polarized intensity.}
\label{fig:Noise}
\end{figure}

\subsection{Source detection and upper limits}
 Not all of the sources detected in total intensity are also detected
 in polarization. In order to fix the flux-threshold in polarization
 to distinguish between detection and upper-limit, we proceeded as
 follows: in the NVSS images we have randomly chosen regions where no
 sources were detected in total intensity. These empty regions have
 been selected to have an area $\geq$ than one beam, to mimic the same
 conditions of the detected sources. We have then computed the mean
 polarized intensity flux, and compared the distribution of
 ``real-sources'' and ``noise-sources'' (see
 Fig. \ref{fig:Noise}). The threshold was fixed at the value of P
 where the number of ``noise-sources'' is less than 10\% than the
 number of ``real-sources''. This ensures that we are safely dominated
 by detections at every P. The threshold corresponds to $P_t=$0.45
 mJy/beam so when $ P \leq P_t$ mJy/beam a 1-$\sigma$ upper-limit for
 $F_P$ was derived: $F_P \leq \sqrt{\frac{P_t^2} {\left( I^2-
     \sigma_I^2\right)}}$.\\ The final sample of sources consists of
 696 sources, 334 of which are upper limits.
\section{Depolarization Analysis}
\label{sec:DP}
The large sample of sources obtained by the NVSS catalogue allows us
to investigate how the depolarization changes as a function of the
projected distance from the cluster center.\\ A radial decline of the
magnetic field proportional to $n^{\eta}$ is expected from
magneto-hydrodynamical simulations performed with different codes
(\citealt{2005xrrc.procE8.10D}, \citealt{1999A&A...348..351D},
\citealt{2005ApJ...631L..21B}, \citealt{2008A&A...482L..13D},
\citealt{2010ApJS..186..308C}) as a result of the compression of
thermal plasma during the cluster gravitational collapse. In addition,
the comparison between thermal and radio profile brightness
\citep{2001A&A...369..441G} also suggests that the non-thermal
component follows the distribution of the thermal gas.  Here we will
assume that the magnetic field radial profile is proportional to the
gas density profile, according to:
\begin{equation}
\label{eq:bprofile}
B(r)=\langle B_0 \rangle\left( \frac{n}{n_0}\right)^{\eta},
\end{equation}
where $\eta$ is a parameter.
Given this assumption, the different values of $r_c$ for the clusters
in our sample have to be accounted for. The gas density
profile (Eq. \ref{eq:Sxbeta}) is almost constant for $r<r_c$ and
decrease faster as $r>r_c$. We have normalized the distance of the
sources to the cluster core radius, defining $r_{norm}=r/r_c$. In the
upper panel of Fig. \ref{fig:DepoAll} the depolarization percentage as
a function of $r_{norm}$ is reported for all the sources and the upper
limits in our sample.
\subsection{Fractional polarization trend}
\begin{figure*}[ht]
\centering
\includegraphics[width=0.7\textwidth]{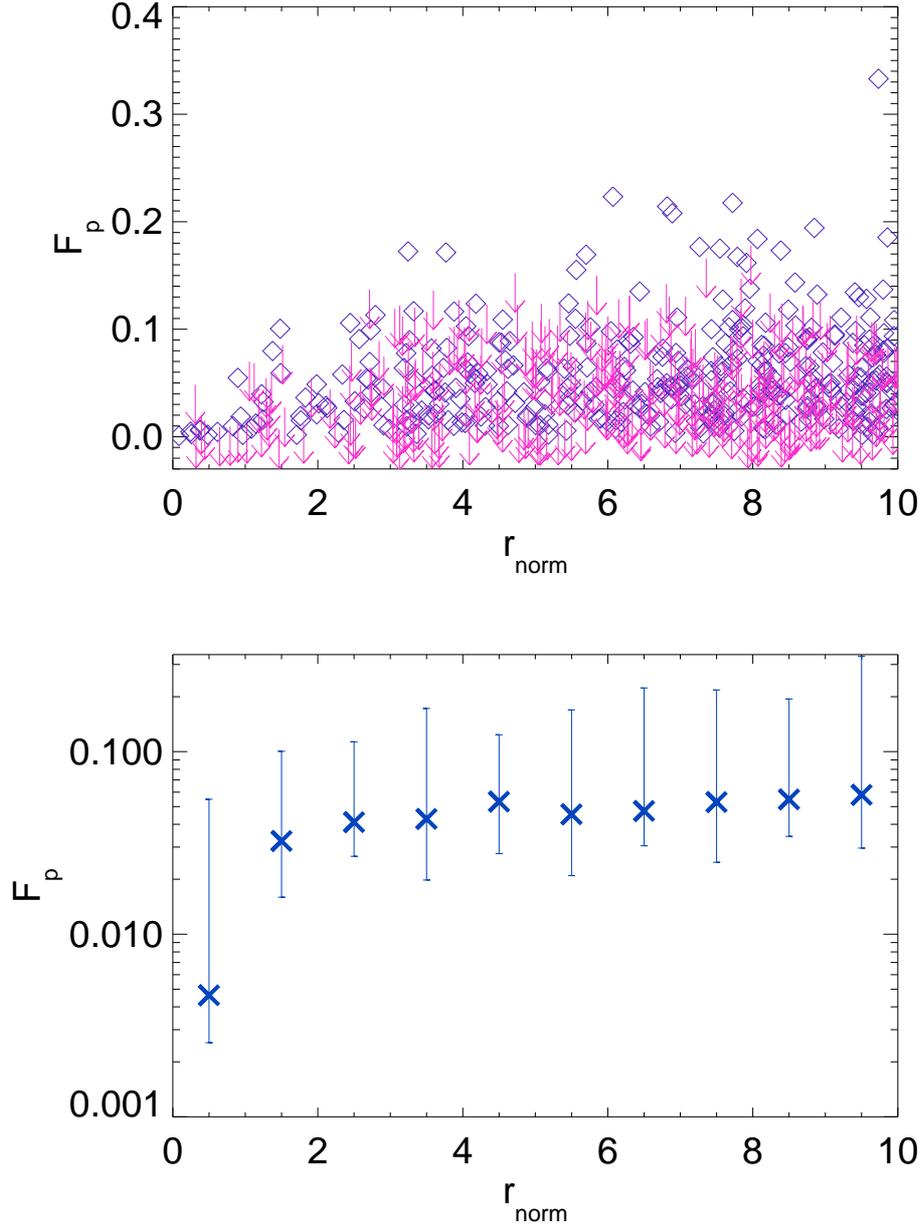}
\caption{Upper panel: Fractional polarization for the sources
  belonging to the cluster sample as a function of the projected
  distance from the cluster center normalized by the core radius. The
  arrows indicate upper limits. Lower panel: the median of the KM
  estimator in every bin is shown, error bars refer to the 16th and
  84th percentile of the KM distribution. }
\label{fig:DepoAll}
\end{figure*}

The sample of sources that we have obtained is affected by the
presence of non-detections, or censored data. Censored data points are
those whose measured properties are not known precisely, but are known
to lie above or below some limiting sensitivity. Here a non-detection
still gives us the useful information that the P flux (and so the
fractional polarization) is below a certain threshold. We have thus a
left-censored sample of data. Our aim is to derive the depolarization
trend of the sample of sources we have selected, taking care of the
information given by censored data. \\An extensive field of statistics
called "survival analysis" of ``lifetime data" exists to address
problems of this kind. A very useful statistical estimator in the
survival analysis is the Kaplan Meier (KM) estimator, that is
described by \citet{1985ApJ...293..192F}. It is a non parametric
maximum likelihood type estimator of the distribution function $F(t)$,
with $t$ being a generic variable, in our case $F_P$. It is usually
expressed in term of the survival function $S(t)$. Let
$x_1<x_2<....<x_r$ denote the distinct, ordered, observed values, the
KM estimator is given by\footnote{Here the KM estimator is expressed
  in the case of right-censored sample. To obtain an estimator of
  $F(t)$ for left-censored samples we have computed $P(T \leq
  t)=S_{KM}(M-t))$, with M being the maximum of $t_i$.}:
\begin{eqnarray}
S_{KM}(t)=Prob(T \geq t)=1-F(t)\\
=\Pi_{i, x_i < t}\left(1-\frac{d_i}{n_i}\right)^{\delta_i}, \mbox{ when } t > x_1\\
=1, \mbox{ when } t \leq x_1
\end{eqnarray}
with $n_i$ being the number of objects (detected or undetected) $\geq
x_i$, $d_i$ are the number of objects at value $x_i$, and
$\delta_i=1,0$ if $x_i$ is detected or undetected
respectively. \\ Using the KM estimator, we have derived the
distribution function of our data set in each bin. In the lower panel
of Fig. \ref{fig:DepoAll} the median of the distribution function in
each bin is reported. The error bars refer instead to the 16th and
84th percentile of the distribution (those that include 68\% of the
data in the distribution).\\ A trend is detected in the $F_P$ going
from the cluster center to the cluster outskirts
(Fig. \ref{fig:DepoAll}). The value of $F_P$ in the first bin,
i.e. for $r_{norm}<$1 is 0.0046, and increases gradually in the outer
bins. The effect of the cluster on the observed $F_P$ is clear out to
$\sim$5 core radii, while in the outer bins, the mean $F_P$ is
scattered around a constant value.  It reaches values going from 0.047
to 0.058 in the 7th to 10th bin. As the distance from the cluster
center increases, $RM$ decreases and the observed $F_P$ is no longer
affected by the presence of the ICM.\\The value that we find in the
outer bin is interpreted as due to depolarization intrinsic to the
structure of the radio source as seen by the radio-telescope at this
resolution and frequency. We assume
that there is no bias in the intrinsic depolarization in all the
sources in our sample. The intrinsic fractional polarization is
assumed to be the same for sources regardless of their projected
distance from the cluster center, as found e.g. in the Coma cluster
\citep{2010A&A...513A..30B}.\\ It follows that internal depolarization
should not affect the observed trend of $F_p$ versus $r_{norm}$, but
rather act like a constant normalization factor.\\ The observed trend
indicates that magnetic fields are common constituents of galaxy
clusters, in agreement with the results by \citet{2004JKAS...37..337C}
and \citet{2004mim..proc...13J}, who analyzed the Faraday Rotation
Measures of sources located behind and within clusters.

\section{Magnetic fields properties}
\label{sec:Bprop}
The observed trend of the fractional polarization $F_P$ versus the
projected distance from the cluster center can be used to constrain
the magnetic field properties. Both the strength and the morphology
({\it i. e.} the power spectrum) of the magnetic fields affect the
Faraday $RM$ of radio sources, and the consequent beam depolarization
that we are investigating here.\\ We use 3D simulations of random
magnetic fields, performed with FARADAY code
\citep{2004A&A...424..429M} to simulate the expected trend of $F_P$
for different magnetic field configurations. In this Sec. the magnetic
field modeling is briefly presented, and the comparison with the
observed sample is performed.
\subsection{Magnetic field modeling}
The FARADAY code \citep{2004A&A...424..429M} simulates 3D models for
the magnetic field in galaxy clusters, and computes the expected $RM$
for radio sources within/behind the cluster.\\ Numerical simulations
start by considering a power-law power spectrum for the vector
potential ${\bf A}$ in the Fourier domain:
\begin{equation}
|A_k|^2 \propto k^{-\zeta}
\end{equation}
 and extract random values of its amplitude $A$ and phase $\phi$. The
 amplitude   $A$ is randomly extracted from a Rayleigh distribution (in
 order to obtain a Gaussian distribution for the real magnetic field
 components), while $\phi$ varies randomly from 0 to 2$\pi$. Here $k$
 is the wave vector in the Fourier domain. The correspondent quantity
 in the real space is $\Lambda = \frac{2 \pi}{k}$.  The magnetic field
 components in the Fourier space are obtained by computing the
 cross-product:
\begin{equation}
{\tilde B}(k) =ik\times {\tilde A}(k).
\end{equation}
Finally, the field components $B_i$ in the real space are derived
using a 3D Fast Fourier Transform inversion. The numerical magnetic
field model is then divergence-free, Gaussian random, and
isotropic. The power spectrum normalization is set such that the
average magnetic field strength follows Eq. \ref{eq:bprofile}. This
operation is performed in the real space domain, to reduce the
computational burden. The same approach has already been used in
several works (\citealt{2004A&A...424..429M},
\citealt{2008MNRAS.391..521L}, \citealt{2008A&A...483..699G},
\citealt{2010A&A...513A..30B}, \citealt{2010A&A...514A..71V}).  \\ The
adopted power spectrum power law introduces three free parameters:
$n$, $\Lambda_{min}$ and $\Lambda_{max}$. Two more free parameters are
introduced once the magnetic field profile (Eq. \ref{eq:bprofile}) is
considered: $\langle B_0 \rangle$ and $\eta$. Two main degeneracies
affect these parameters: the first one concerns the power spectrum,
and in particular the values of $n$ and $\Lambda_{max}$, and another
degeneracy affects $\langle B_0 \rangle$ and $\eta$. Different
combinations of $n$ and $\Lambda_{max}$ as well as $\langle B_0
\rangle$ and $\eta$ lead to similar values of fractional polarization,
since it depends mainly on the amount of Faraday Rotation Measures
that originates in the ICM causing the beam depolarization.\\ Given
the number of free parameters, we have fixed the magnetic field radial
profile slope ($\eta$), and for three different magnetic field power
spectra we have investigated the magnetic field strength at the
cluster center.\\ The magnetic field power spectrum is assumed to be
Kolmogorov-like. Different independent analyses of $RM$ data
(\citealt{2005A&A...434...67V}, \citealt{2008A&A...483..699G},
\citealt{2010A&A...513A..30B}) indicate that the magnetic field power
spectrum observed in galaxy clusters is in good agreement with the
power law expected by the Kolmogorov theory, which in our 3D notation
corresponds to $n=11/3$. We note that the theory developed by
Kolmogorov is applicable incompressible un-magnetized and uniform
fluids, so that its application to the case of the ICM of galaxy
clusters is all but obvious. Nonetheless, observational data and
cosmological simulations indicate that it is a good description of the
pseudo-pressure fluctuations \citep{2004A&A...426..387S}, velocity
field \citep{2009A&A...504...33V} and magnetic field (
\citealt{2005A&A...434...67V}, \citealt{2008A&A...483..699G},
\citealt{2010A&A...513A..30B}) in the ICM. Although other spectral
slope are not excluded by data, we will adopt here a Kolmogorov like
power spectrum. The power spectrum maximum and minimal scale were set
to reproduce a magnetic field power spectrum having an
auto-correlation length $\Lambda_B\sim$164 kpc (model 1),
$\Lambda_B\sim$ 25 kpc (model 2) and $\Lambda_B\sim$ 8 kpc (model
3). We then adopted $\Lambda_{min}=$8 kpc, much smaller than the beam
linear size. The values of $\Lambda_{max}$ are then 512 kpc (model 1),
64 kpc (model 2), and 14 kpc (model 3). The value of $\eta$ is set to
0.5. This choice is motivated by recent works
(\citealt{2008A&A...483..699G}, \citealt{2010A&A...513A..30B}) that
found the best agreement between simulations and observations when the
magnetic field energy density follows the thermal gas density ({\it}
i. e. $\eta$=0.5).  We then investigated different values of $\langle
B_0 \rangle$=1, 5, and 10 $\mu$G.

\subsection{Comparison with observed data}
\label{sec:simVsData}
\begin{figure}[ht]
\centering
\subfigure{\includegraphics[width=\columnwidth]{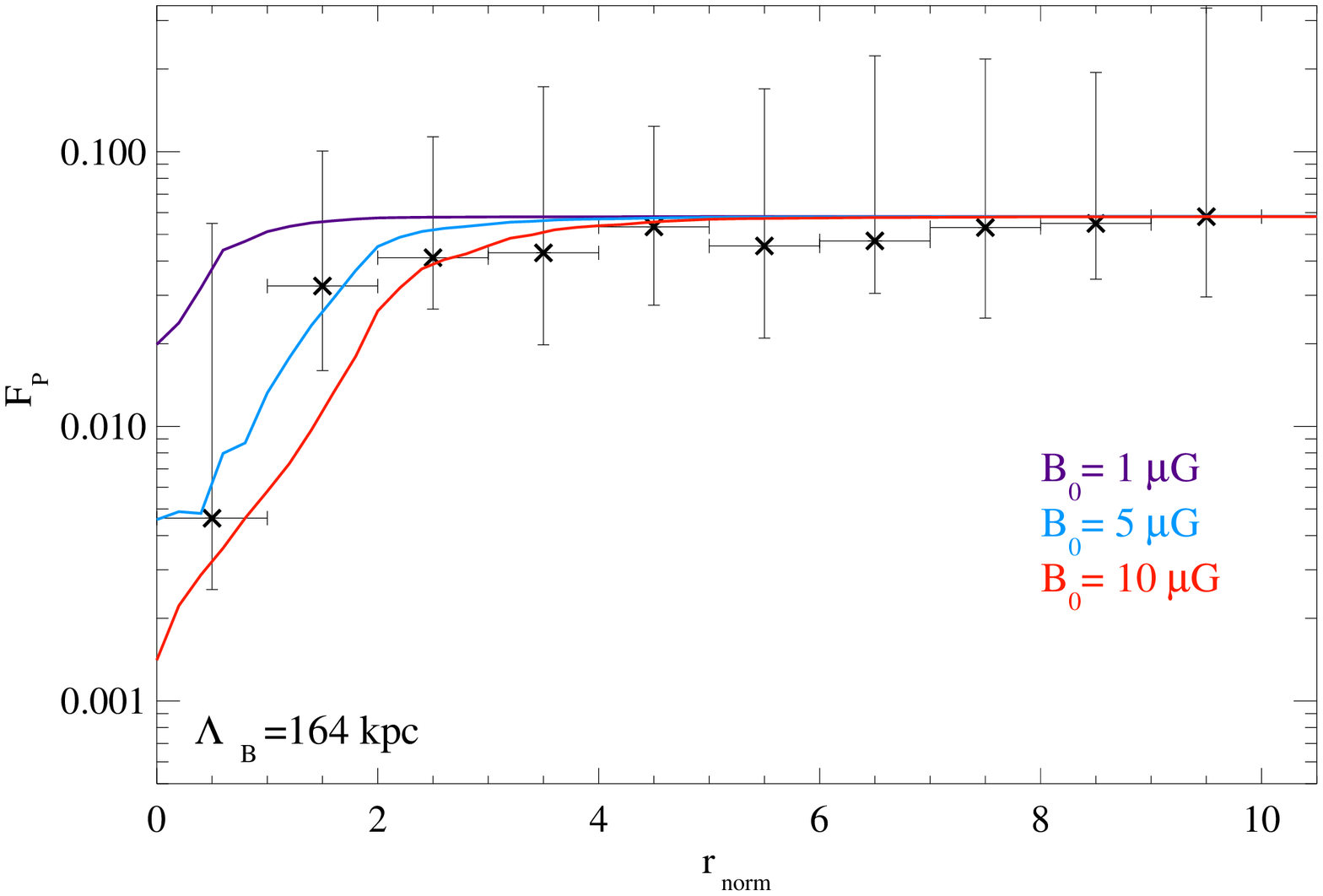}}
\subfigure{\includegraphics[width=\columnwidth]{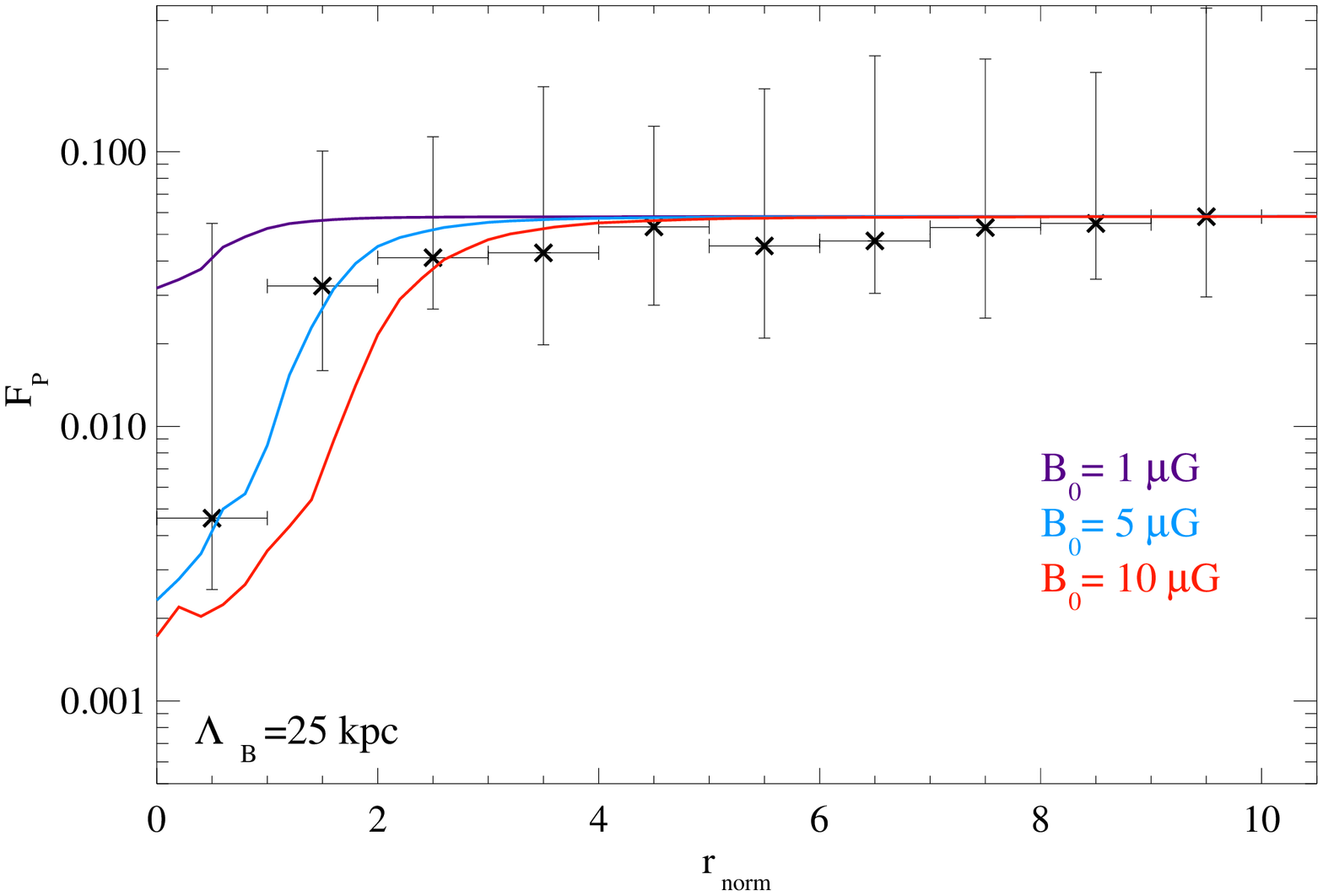}}
\subfigure{\includegraphics[width=\columnwidth]{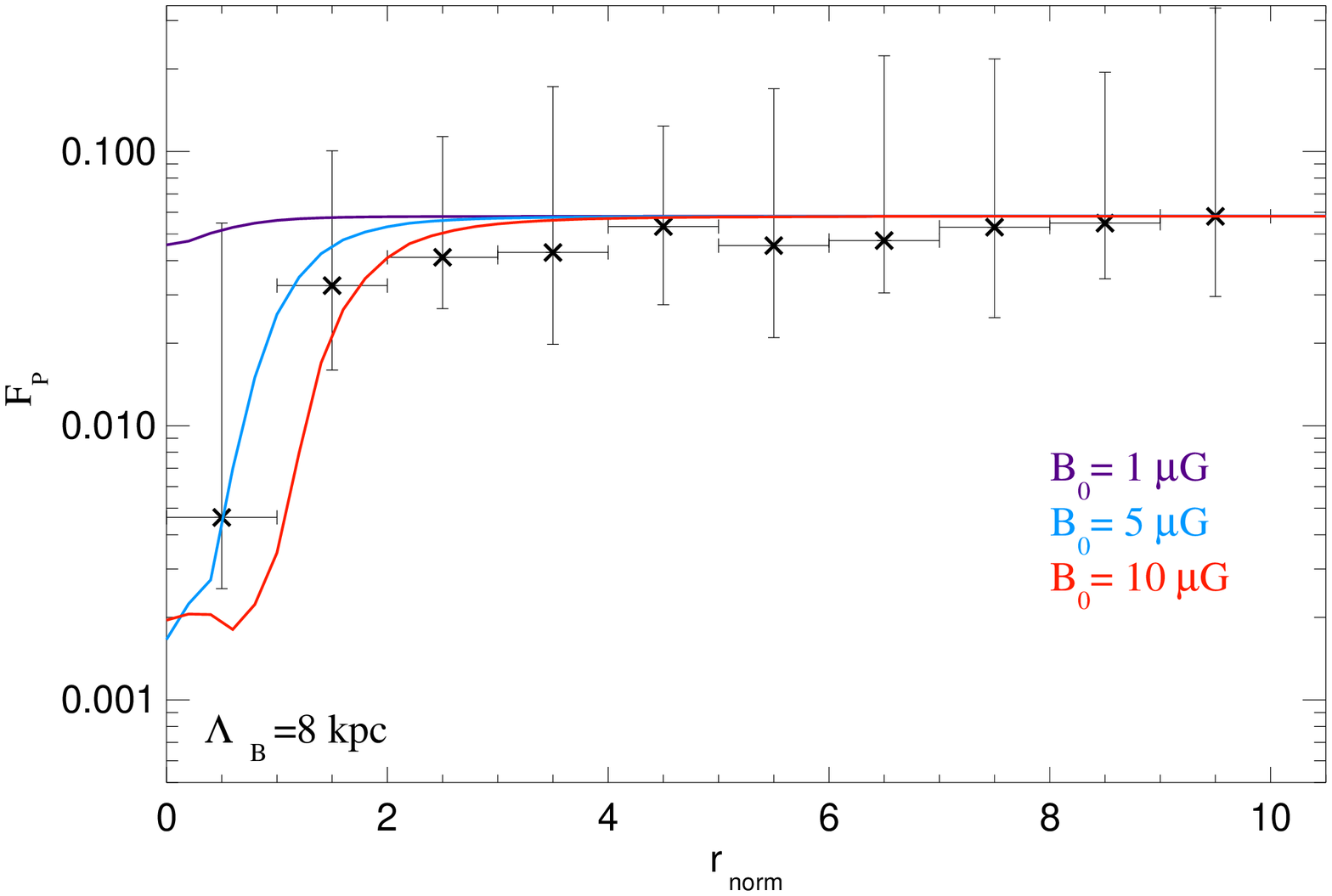}}
\caption{Simulated (lines) and observed (crosses) depolarization
  trends for models with different $\Lambda_B=$164,25 and 8 kpc (top,
  middle and bottom panel respectively). Different colors refer to
  different magnetic field strengths, as reported in the figures.}
\label{fig:DPsim}
\end{figure}

Numerical simulations of the magnetic fields have been performed in a
cubical box of 1024$^3$ pixels, with a pixels-size of 2 kpc. The
simulated magnetic field is periodic at the grid boundaries, and the
computational grid has been replicated to obtain a simulated field of
view of $\sim$ 20 $\langle r_c\rangle \times$ 20 $\langle r_C
\rangle$, with $\langle r_C \rangle$ being the mean value of $r_c$ for
the clusters in the sample weighted by the number of sources in each
cluster ($\sim$ 270 kpc).\\ Following \citet{2004A&A...424..429M} our
aim is to compute, for different magnetic field models, the ratio
between the polarization obtained after the rotation of the plane of
polarization and the intrinsic polarization at a given frequency. This
is the quantity that can be compared with the observed $F_P$, once the
effects introduced by observations are considered.\\ 
We simulated different magnetic field models and derived
the expected $F_P$ for sources located halfway through the cluster at
increasing distance from the cluster center. We proceeded as follows:
\begin{itemize}
\item We simulated different magnetic field models, with $\langle B_0
  \rangle=$ 1, 5 and 10 $\mu$G, and $\Lambda_B=$ 8, 25, and 164 kpc.
\item The intrinsic polarization $P_{int}$ of the sources is assumed
  to be spatially uniform and ordered ($\Psi_{int}$ constant) and its
  value is set to reproduce the observed value of $F_P $ at
  $r_{norm}=$10.
  \item According to Eq. \ref{eq:psiobs}, we derived the synthetic $RM$
    images expected for the different magnetic strengths. The
    integration path goes from 0 (cluster center) to 10 $\langle r_c
    \rangle$. We used $n_0=0.004$ cm$^{-3}$, $\beta=0.67$ that are the
    mean values of our sample weighted by the number of sources that
    are in each cluster.  We produced simulated $RM$ images on a
    5400$\times$5400 kpc field of view, with a resolution of $\sim$2
    kpc.
\item We then computed the expected polarization plane direction
  according to Eq. 1 and taking into account the effect
  due to the NVSS finite bandwidth.
\item $P_{int}$ and $\Psi_{obs}$ were then converted into the Stokes
  parameters $Q(\lambda)$ and $U(\lambda)$ following Eq. \ref{eq:UQ}.
\item To the simulated images of $Q(\lambda)$ and $U(\lambda)$ we
  added a Gaussian noise having rms$=$0.3 mJy/beam, as the mean NVSS
  noise for $Q$ and $U$ images. $Q$ and $U$ images were then convolved
  with a Gaussian beam having major and minor axis $\approx$ 42
  kpc. This is the linear size corresponding to 45$''$ at z$=$0.048
  that is the median redshift of the cluster sample weighted by the
  number of sources in each cluster.
\item Finally, the synthetic $Q$ and $U$ images were transformed back
  to $P$ using Eq. \ref{eq:UQ}.
\item The trend of simulated and observed $P$ versus $r_{norm}$ have
  been compared.
\end{itemize}
The comparison between the simulated and observed $F_P$ trend obtained
by averaging data from the clusters in the sample is performed under
the following assumptions:
\begin{enumerate}%[i] 
\item the magnetic field is assumed to be a Gaussian random
  field. This can be incorrect if the magnetic field is intermittent,
  and characterized by filaments. However, this is a common assumption
  in this field, and is mainly motivated by the fact that current
  data are generally too sparse to determine higher order correlation
  (see e.g. \citealt{2008MNRAS.391..521L}). \\
\item The magnetic field phases are assumed to be random, which means
  that the magnetic field is assumed to have no preferred
  orientation. Again, this may be not the case when individual
  clusters are analyzed ( e. g. M 84,
  \citealt{1987MNRAS.228..557L}) but we are interested in the magnetic
  field properties averaged over large volumes of several clusters.\\
\item The power spectrum fluctuations are assumed to vary with the
  thermal gas density in the ICM, according to
  Eq. \ref{eq:bprofile}.\\
\end{enumerate}
Under these assumptions, the model that best reproduces the observed
trend of $F_P$, among those considered here, has $\langle B_0
\rangle=5 \mu$G (Fig. \ref{fig:DPsim}).\\ The simulated $F_P$
trends for the different magnetic field models are shown in
Fig. \ref{fig:DPsim}. We are aware that we are probably
averaging the $F_P$ produced by slightly different magnetic field
models and configurations, and that more refined and cluster specific
models should be considered if precise values are needed. Nonetheless,
the best-fit model gives a reasonable order of magnitude estimate for
the average magnetic field properties in galaxy clusters. 

In our best fit model with a central magnetic field $\langle B_0
\rangle=5 \mu$G, the mean magnetic field over the central Mpc cube,
$B_{\langle 1 Mpc \rangle}$, resulting from our best fit model is
$\sim$2.6 $\mu$G.  A more detailed analysis has been performed for
some of the clusters in our sample: Abell 2255
(\citealt{2006A&A...460..425G}), Hydra-A
(\citealt{2005A&A...434...67V}, \citealt{2008MNRAS.391..521L}) and
Coma (\citealt{2010A&A...513A..30B}). Given the degeneracy between
$\langle B_0 \rangle$ and $\eta$ we compare the mean magnetic field
over the central Mpc cube, found by the above mentioned authors with
the one obtained here. In A2255 and in the Coma cluster
\citet{2006A&A...460..425G} and \citet{2010A&A...513A..30B} find
$B_{\langle 1 \, Mpc \rangle}\sim$1.2 $\mu$G, and $B_{\langle 1 \, Mpc
  \rangle}\sim$2 $\mu$G respectively, while the analysis by
\citet{2005A&A...434...67V} yields $B_{\langle 1 \, Mpc \rangle}\sim$1
$\mu$G for Hydra A. They are compatible with the value obtained here.

\begin{figure}[ht]
\centering  
\includegraphics[width=0.45\textwidth]{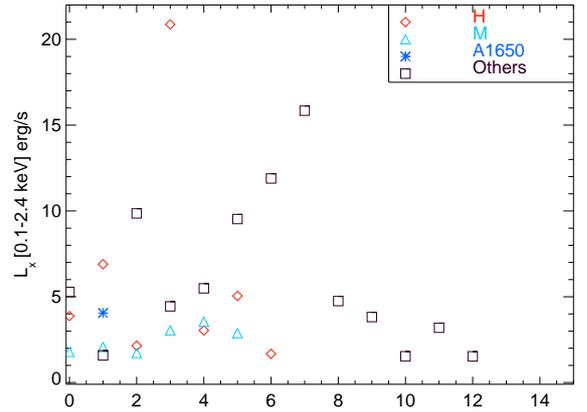}
\caption{X-ray luminosity in the 0.1-2.4 keV band for the clusters in
  the sample here analyzed. Different symbols and colors refer to
  different sub-samples. Clusters that host a radio halo (red
  diamonds); clusters where a radio halo has not been detected so far,
  but with signature of on-going merger according to the literature
  (cyan triangles), other clusters (squares). The cluster A1650, whose
  merging state is debated, is signed with a blue asterisk.}
\label{fig:lxHalo}
\end{figure}

\begin{figure*}[ht]
\centering
\subfigure{\includegraphics[width=0.45\textwidth]{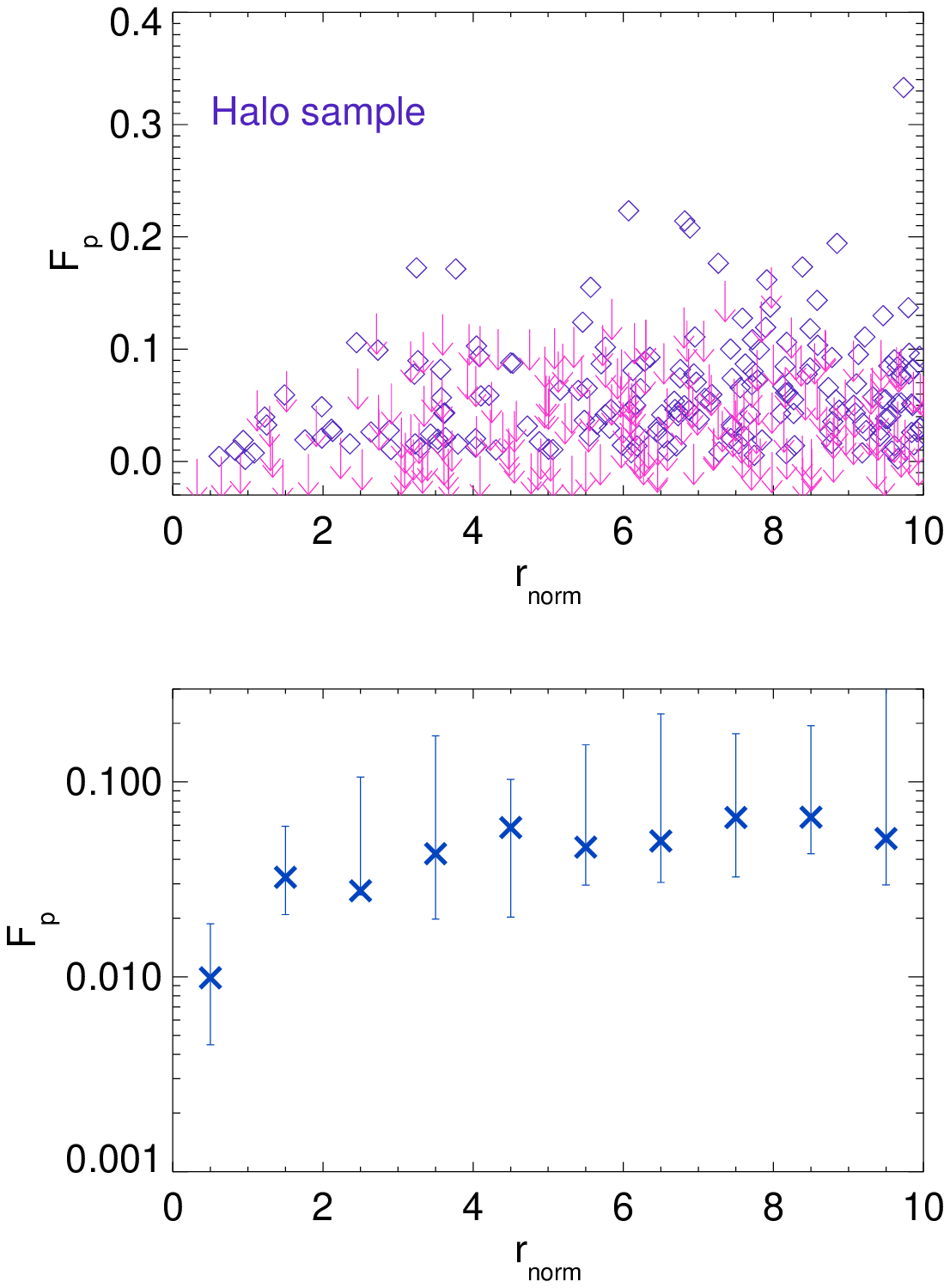}}
\subfigure{\includegraphics[width=0.45\textwidth]{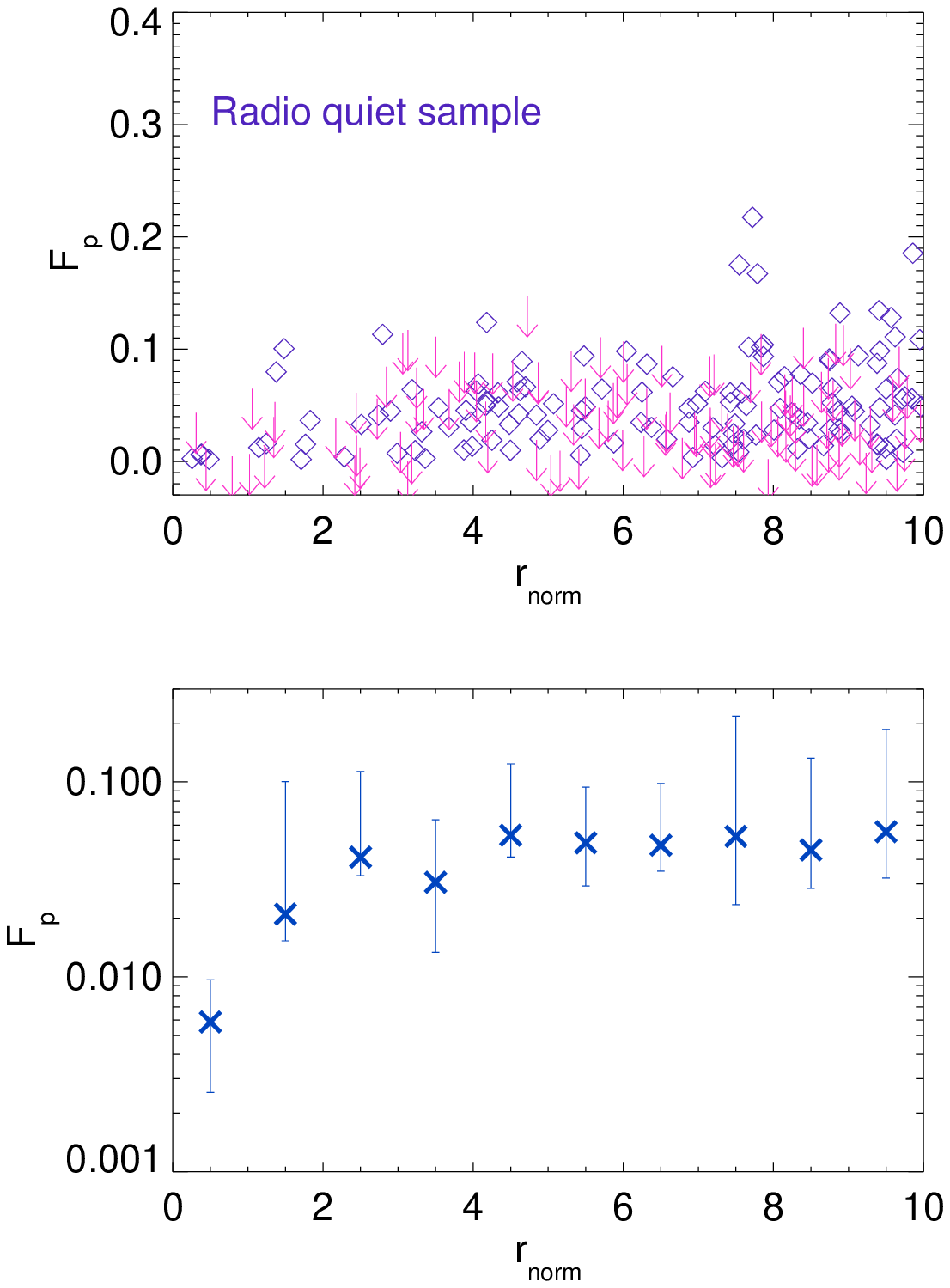}}
\caption{ Fractional polarization versus the projected distance from
  the cluster center, normalized by the core radius, for sources in
  clusters that host a radio halo (left), and that do not host a radio
  halo (right). Arrows indicate upper limits.  Bottom panel: crosses
  refer to the median of the KM estimator is each bin, bars indicate
  the 14th and 84th percentile. Due to the low number of points in
  each bin the width of each bin is adaptively set so that at leat 3
  detections fall within a bin. }
\label{fig:DepoHalo}
\end{figure*}

\begin{figure*}[ht]
\centering
\subfigure{\includegraphics[width=0.45\textwidth]{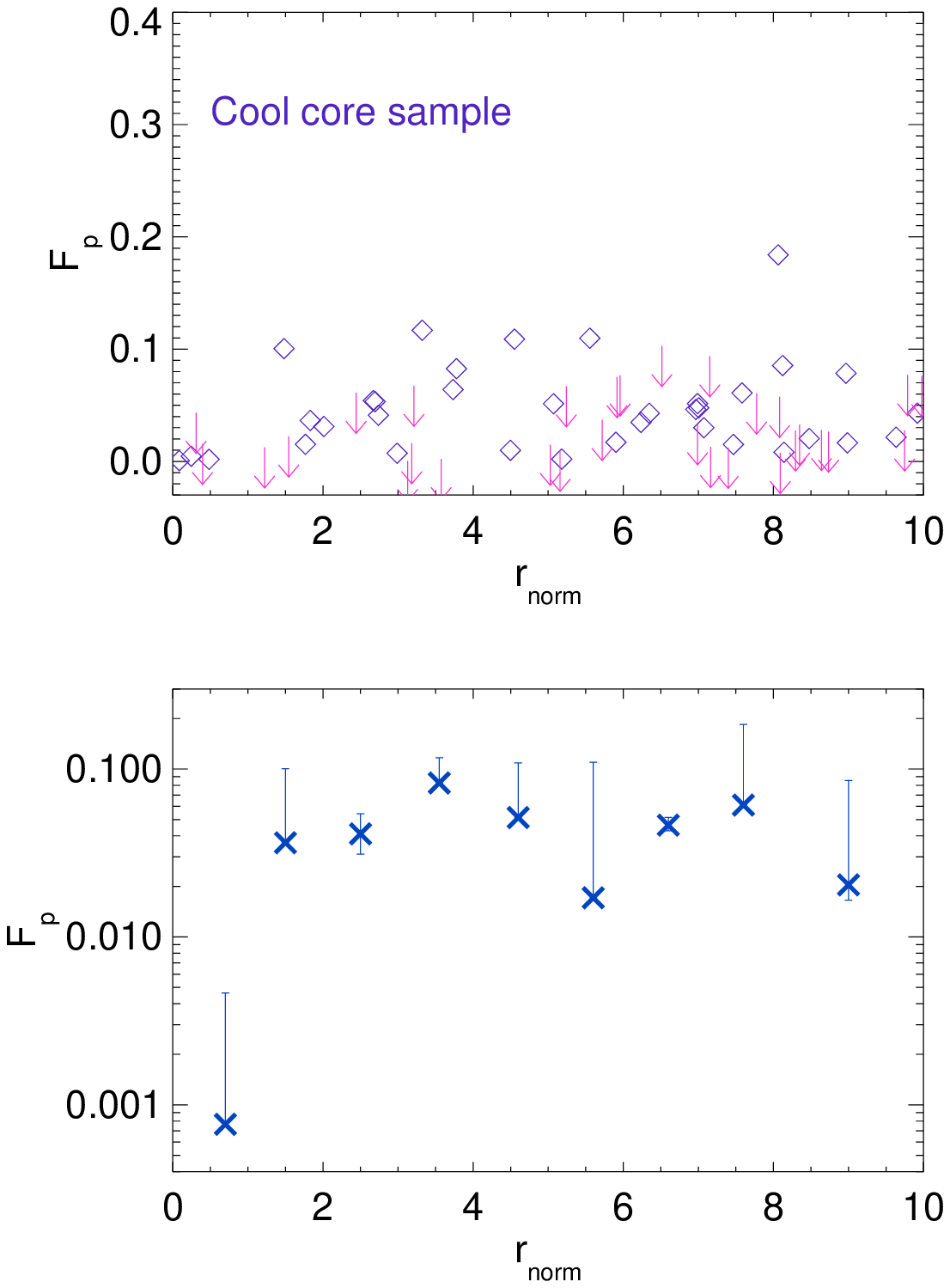}}
\subfigure{\includegraphics[width=0.45\textwidth]{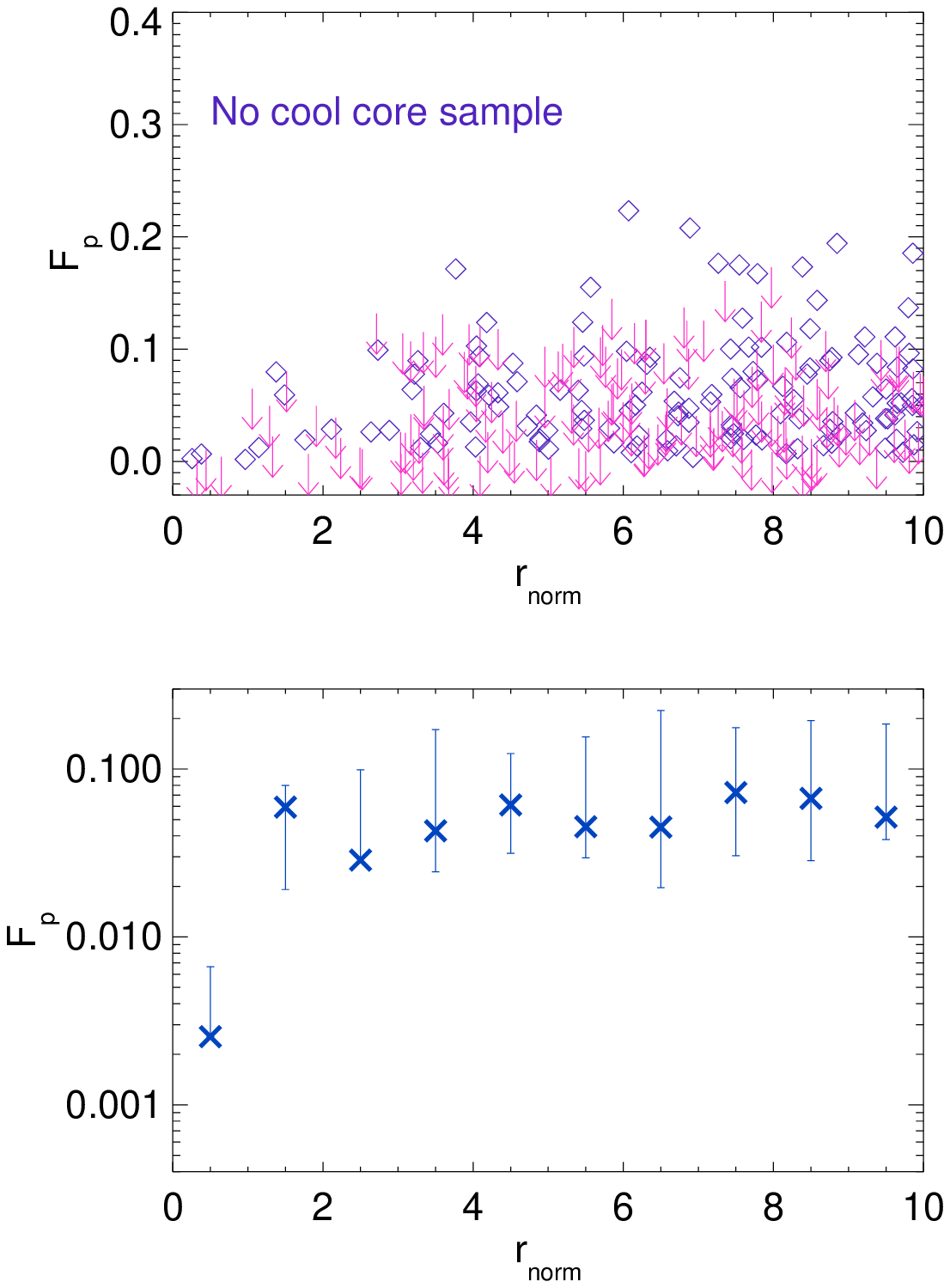}}
\caption{Top panel: Fractional polarization versus the projected
  distance from the cluster center, normalized by the core radius, for
  sources in cool core clusters (left) and non cool core clusters
  (right). Arrows indicate upper limits. Bottom panel: crosses refer
  to the median of the KM estimator is each bin, bars indicate the
  14th and 84th percentile. Due to the low number of points in
  each bin the width of each bin is adaptively set so that at leat 3
  detections fall within a bin. }
\label{fig:DepoCC}
\end{figure*}

\begin{figure*}[ht]
\centering
\subfigure{\includegraphics[width=0.45\textwidth]{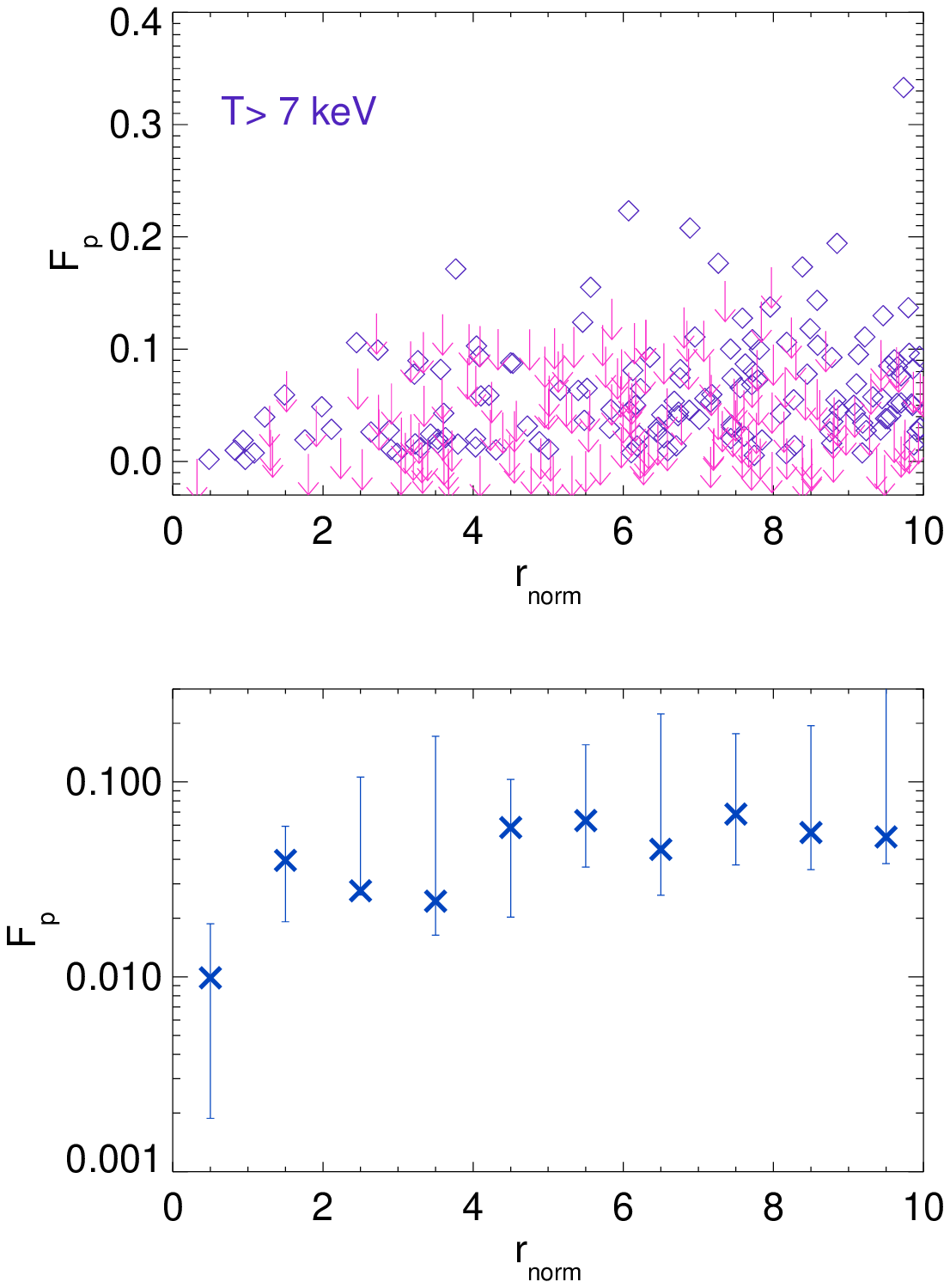}}
\subfigure{\includegraphics[width=0.45\textwidth]{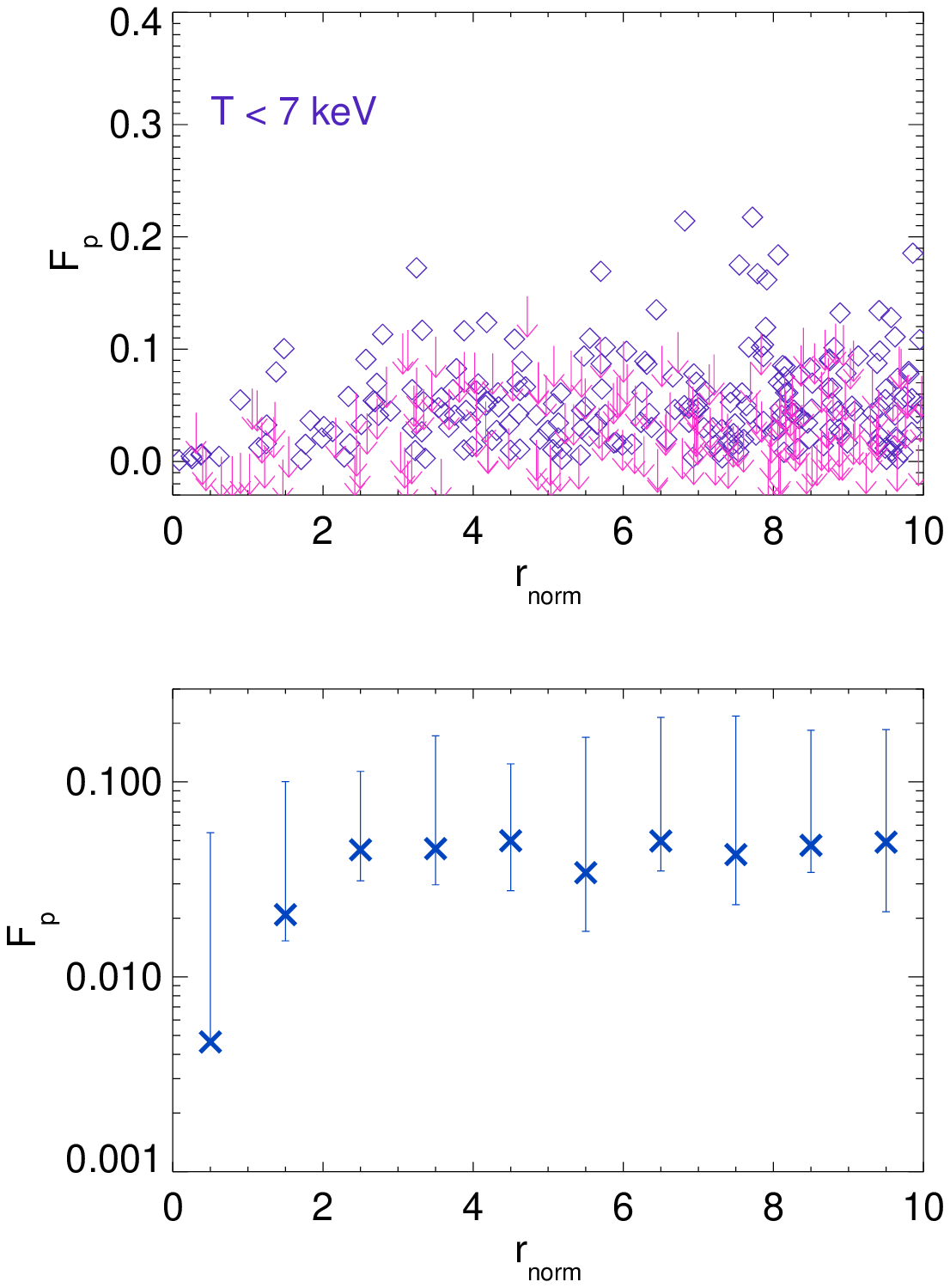}}
\caption{ Fractional polarization versus the projected distance from
  the cluster center, normalized by the core radius, for sources in
  clusters with T$\geq$ 7 keV (left), and clusters with T$\leq$ 7 keV
  (right). Arrows indicate upper limits. Bottom panel: crosses refer
  to the median of the KM estimator is each bin, bars indicate the
  14th and 84th percentile. Due to the low number of points in
  each bin the width of each bin is adaptively set so that at leat 3
  detections fall within a bin. }
\label{fig:DepoT_fede}
\end{figure*}

\section{Dependence of magnetic field properties in clusters with and without radio halo}
\label{sec:halo}
The sample of clusters that we have selected comprises both virialized
and merging systems (see Tab. \ref{tab:sample2} and references
therein). In a fraction of clusters, with signatures of recent or
on-going mergers, a radio halo has been detected (see e.g.
\citealt{2008A&A...484..327V}, \citealt{2009A&A...507.1257G}).
The origin of these radio sources is still un-known, and two main
classes of models have been proposed in the literature:
\begin{itemize}
\item {\it primary or re-acceleration models}: in which electrons are
  re-accelerated {\it in situ} through second-order Fermi mechanism by
  ICM turbulence developing during cluster mergers
  (e.g. \citealt{2001MNRAS.320..365B}; \citealt{2001ApJ...557..560P});
\item {\it secondary or hadronic models}: in which electrons originate
  from hadronic collisions between the long-living relativistic
  protons in the ICM and thermal ions (e.g. \citealt{1980ApJ...239L..93D}).
\end{itemize}
One possibility to explain the presence of radio halos in only a
fraction of clusters is to assume different magnetic field strength in
the ICM of clusters with and without radio emission. The magnetic
field could be amplified during cluster mergers, as indicated by MHD
cosmological simulations (see review by
\citealt{2008SSRv..134..311D}), and then, it could potentially be
dissipated once the cluster turns back to the equilibrium
state.\\ 
\subsection {Clusters with and without radio halo}
With the aim of investigating a possible difference between the
magnetic field properties in clusters with and without radio halos, we
have divided our initial sample in two sub-samples, that contain
clusters with radio halos (labeled with H in Tab. \ref{tab:sample2})
and clusters where a giant radio halo (i.e. with size $\sim$1 Mpc) has
not been detected so far.   It is worth noting that no upper
  limits have been put on the radio emission of clusters where a radio
  halo has not been detected, so that additional care is required to
  avoid, or at least limit, obvious observational bias.\\ Most of
  radio halos have been discovered with follow-up observations of NVSS
  candidates. Since the NVSS cannot detect large (Mpc) scale emission
  in clusters at z$<$0.044, we exclude from our initial sample the
  clusters with such a redshift. Note that this is a conservative
  approach, since other single-dish instruments would be clearly able
  to detect radio emission in these very nearby objects, as in the
  case of the Coma cluster.\\ In addition, the radio power of the
  observed radio halos is usually found to correlate with the cluster
  X-ray luminosity (e.g. \citealt{2000ApJ...544..686L},
  \citealt{2000cucg.confE..37F}). Our sample includes clusters with
  X-ray luminosities going from 1.5 to 20.8 $\times$10$^{44}$
  erg/s. Clusters with the lower X-ray luminosity could host halos of
  lower radio power, that are undetected by present surveys (see e.g.
  \citealt{2005ASPC..345..227C}). It is worth mentioning that
  recently two radio halos have been found in clusters with low X-ray
  luminosity, namely 0217+070 \citep{2011ApJ...727L..25B}, and A1213
  \citep{2009A&A...507.1257G}, whose radio power is an order of
  magnitude or more above the extrapolation of the $L_X-P_R$
  correlation at low X-ray luminosities. However, we decided to adopt
  a conservative approach, and to compare clusters in the same X-ray
  luminosity range. In Fig. \ref{fig:lxHalo}, the X-ray
  luminosity of the clusters in the energy range 0.1- 2.4 keV is
  shown. Clusters are represented with different colors and symbols,
  depending on the presence of radio halo and on their dynamical
  state. Objects with z$<$0.044 have been removed. Clusters with and
  without radio halo populate the $L_X$ range uniformly.
%According to the $L_X-P_{1.4
%    GHz}$ correlation, the cluster with lower X-ray luminosity in our
%  sample could host a radio halo with $P(1.4 GHz) \sim 1.4
%  \times 10^{23}$W/Hz.  This power translates into a radio brightess of

Among the clusters that are known to host a radio halo, the cluster
A3562 is the one with the lowest X-ray luminosity: $1.67x10^{44}$
erg/s. The radio halo was discovered by \citet{2000MNRAS.314..594V}
by inspecting NVSS images.  We will then consider in the following
comparison only clusters with $L_X$ higher than the one of
A3562.\\ Although these two additional criteria should prevent us from
obvious observational bias, it is worth mentioning that radio halos
with very steep radio spectra and/or low radio brightness could still
be present in the cluster labeled as ``radio--quiet''. Only future
instruments, such as LOFAR, will be able to provide a definitive
answer. Nonetheless, it is interesting to investigate the properties
of the ICM magnetic fields in these two sub-samples, according to our
present knowledge about their radio emission.

  The fractional polarization for the two sub-samples
(radio--halo and radio--quiet) is shown in Fig. \ref{fig:DepoHalo}. The
halo sample consists of 374 sources (194 of which are upper limits),
while the radio--quiet sample consists of 243 sources (112 upper
limits).\\ We want to test the null hypothesis that the two
populations, subject to left-censoring, have the same distribution.
We performed the logrank statistical test, that is a non-parametric
test, widely used in astronomy when censored populations have to be
compared. \\ According to the logrank test, let $T_{ij}$ being the
observed value from distribution $i=1,2$, with $j=1...N_i$. Let then
$y_1<y_2<.......<y_r$ with $r \leq N=N_1+N_2$ denote the uncensored
values in the combined samples.  Given two independent random samples
drawn from populations having distribution functions\footnote{Here the
  logrank test is expressed in the case of right-censored sample.}
$F_i(t)=P(T_{ij}\geq t)$ we want to test the null hypotheses $H_0:
F_1(t)=F_2(t)$, for all t. Let $d_{i,j}$ denote the number of objects
from sample $i$ $=y_j$, $d_{j}$ the number of objects from both
samples $=y_j$, $n_{i,j}$ the number of objects from sample $i$ $\geq
y_j$ and similarly $n_j $ the number of objects from both samples
$\geq y_j$. The quantity
\begin{equation}
L_n=\Sigma_{j=1}^{r}\left(d_{1,j}-\frac{d_jn_1}{nj}\right)
\end{equation}
for large $N$, under $H_0$, is approximately normally distributed with
zero mean and variance $\sigma_n^2$.   Hence, $H_0$ is rejected at level $\alpha$ if
\begin{equation}
\left|\frac{L_n}{\sigma_n}\right|\geq z_{\alpha/2}
\end{equation}
where $z_{\alpha/2}$ is the score such that the area under a standard
normal curve over the interval $[-z_{\alpha/2},z_{\alpha/2}]$ is
1-$\alpha$. We refer again to \citet{1985ApJ...293..192F} for an
illustrative explanation of the test with application to astronomical
data. Our null hypothesis is that the two
populations of $F_P$ in clusters with and without radio halo are
different realization of the same sample.
The null hypothesis is accepted with high significance: $P=$ 0.99. We can
conclude that the depolarization trend for sources seen through
clusters is the same in these two sub-samples.
Since the amount of depolarization depends on the Faraday rotation in the
external screen, there are two different possibilities to explain
our results:
\begin{enumerate}
\item{Both magnetic field and gas density distributions
 in cluster with and without radio halos have similar properties, or}
\item{magnetic field and gas density are both different in clusters
  with and without radio halo, but conspire to give rise to the
  same $F_P$. (i.e. higher magnetic field and lower gas density in one
  sample and lower magnetic field and higher gas density in the other
  one)}
\end{enumerate}
The mean values of $n_0$, weighted by the number of sources, for the
radio halo and radio quiet samples are 0.003$\pm$0.001 cm$^{-3}$ and
0.005$\pm$0.010 cm$^{-3}$ respectively. Since they are compatible
within errors, we can conclude that there is no evidence for a
different magnetic field in clusters with and without radio halos.
\subsection{Analysis of the possible contamination by merging effects}
Processes related to merger events could change the magnetic field
structure in galaxy clusters, increasing the field auto-correlation
length and thus also the $F_P$ of the observed sources. Also, if
merger events amplify the magnetic field strength, a lower value of
$F_P$ is expected. These two effects could conspire to give the same
$F_P$ distribution in the radio halo and radio quiet samples even if
magnetic field is higher in the radio halo sample, because of the
larger auto-correlation length of the magnetic field. In order to
prevent this, we have repeated the analysis described in
Sec. \ref{sec:halo} considering only merging systems in the radio
quiet sample. The radio quiet merging sample consists of 203 sources
(94 upper limits), and the logrank test yields a probability of 0.70
that the radio halo and radio quiet merging sample are drawn to the
same intrinsic population. We can conclude that the depolarization
trend is the same in merging clusters, regardless of the presence of a
radio halo or not.

\subsection{Magnetic field and radio halo origin}
The results we have obtained in the previous sections can also be used
as a test for the radio halo formation
models. \\ \citet{2009A&A...507..661B} have analyzed the well-known
correlations between Radio halo power at 1.4 GHz, $P_{1.4}$, and the
cluster X-ray luminosity, $L_X$ (e. g. \citealt{2000ApJ...544..686L},
\citealt{2009A&A...507.1257G}) in a sample of X-ray selected
clusters. Among this sample only $\approx \frac{1}{3}$ of clusters
host radio halos, while for the other clusters of the sample upper
limits on their radio emission have been put
\citep{2007ApJ...670L...5B}. It must be noted that the sample of
clusters that they have analyzed is composed of both merging and
dynamically relaxed clusters, and that the upper-limits have been
computed assuming a value for the radio spectral index.
\citet{2009A&A...507..661B} conclude that in the hadronic scenario the
energy density of the magnetic field in clusters without a radio halo
should be at least 10 times lower than in those that host radio halo
emission.\\ More recently, \citet{2010arXiv1003.1133K} have proposed a
model within the hadronic scenario, in which cosmic rays diffuse away
from their sources (supposed to be supernovae), whereas the magnetic
fields are amplified by mergers in clusters with radio halos. The
observed bi-modality in the $P_{1.4}$-$L_X$ plane is then attributed
to a bi-modality in magnetic field strength for clusters with and
without radio halos. Radio halos are associated with clusters having
{\it strong} B, i. e. $B >> B_{CMB}$ in the radio emitting region,
whereas clusters with $ B <<B_{CMB}$ all over their volume would be
radio quiet\footnote{$B_{CMB}= 3.2 (1 +z)^2 \mu$G is the equivalent
  magnetic field strength of the cosmic microwave background
  (CMB).}.\\ The result obtained in the previous sections indicates
that the bi-modality in the $P_{1.4}$-$L_X$ plane cannot be attributed
to a bi-modality in magnetic field properties.\\  The result we
  find here is based on a statistical approach, however, previous
  works have already investigated the magnetic field properties in
  clusters either with and without radio halo, finding results
  consistent with those we obtain here. For example, magnetic
  fields of several $\mu$G were inferred in clusters without radio
  halo by e.g. \citet{2001ApJ...547L.111C}, and
  \citet{2004A&A...424..429M}, and recently,
  \citet{2010A&A...522A.105G} found that clusters seem to follow a
  common $S_X-RM$ distribution, $S_X$ being the X-ray surface
  brightness, independently on their radio properties. However, this
is the first time that a statistical comparison between magnetic
fields in clusters with and without radio emission has been performed
and our result is a further indication that hadronic models are
difficult to reconcile with present data
(\citealt{2008Natur.455..944B} and \citealt{2009ApJ...699.1288D};
\citealt{2010MNRAS.401...47D}; \citealt{2010arXiv1006.1648J}).

\section{Dependence of magnetic field parameters on the presence of a cool core}
\label{sec:coolcore}
Galaxy clusters are often divided, according to their X-ray emission,
into cool core and non cool core clusters
(e.g. \citealt{2010A&A...513A..37H}). The former class is
characterized by a bright central peak in the cluster X-ray surface
brightness, a high central gas density, and a positive gradient of the
metal abundance profile. The core is characterized by a radially
increasing temperature. Cool core clusters are believed to be systems
in dynamical equilibrium, while the energy released by cluster mergers
could transform cool core clusters into non cool core systems (see
e.g. \citealt{2004ApJ...606..635M}, \citealt{2010A&A...510A..83R} and
references therein).  In the core, the cooling time of the gas is much
shorter than a Hubble time, but the lack of strong cooling lines
revealed by X-ray spectroscopy of Chandra and XMM-Newton, indicates
that the cooling of the gas must be inhibited. Several processes could
in principle be responsible for that, such as the presence of heating
sources (e.g. \citealt{1995MNRAS.276..663B}), or thermal conduction
from larger radii, e.g. \citealt{2001ApJ...562L.129N}). Recently,
several authors have investigated the role that magnetic fields could
have in this context (e.g. \citealt{2010ApJ...713.1332R},
\citealt{2010ApJ...712L.194P}).\\ We will compare here the
depolarization trend of sources in clusters with and without cool
core. \citet{2007A&A...466..805C} analyzed the basic properties of the
clusters in the HIFLUGCS sample (see Tables 1 and 2 of that paper) and
divided the objects in moderate cool core, pronounced cool core and
non cool core. Both moderate and pronounced CC are labeled as CC in
our Tab. \ref{tab:sample2}.\\ We have divided our initial sample into
two sub-samples.  Since cool core clusters are characterized by the
presence of a cD galaxy sitting in the cluster center, while the same
is not always true for non cool core cluster clusters, we considered
in the non cool core sample only clusters that have a radio galaxy at
their center (A1736, A2163, A3558, A3562, COMA, and ZWCL1215).  The
cool core sample consists of 66 radio sources (30 of them are upper
limits), while the non cool core samples comprises 295 sources (165
upper limits).  $F_P$ data are shown in Fig. \ref{fig:DepoCC}.  We
want to test the null hypothesis that the two populations, subject to
left-censoring, have the same distribution.  We performed the logrank
statistical test, The result of the logrank tests applied to the
samples of cool core and non cool core clusters is that the null
hypothesis has a small significance (P=0.15).  Although the null
hypothesis cannot be rejected at 95\% confidence level (corresponding
to P=0.05), that is the threshold commonly assumed in statistics, the
low value of P indicates that the two samples are likely to be
realizations of intrinsically different populations. This could be due
to a different magnetic field or to the different gas density that
characterize the sample (we find a weighted mean of
$n_0=$0.017$\pm$0.013 for the cool core sample and
$n_0=$0.003$\pm$0.001 for the non cool core sample), or a combination
of the two.\\
Increasing attention has been devoted in recent years to the role that
different magnetic field configurations could have in explaining the
observed bi-modality among cool core and non-cool core clusters (see
e.g. \citealt{2010ApJ...712L.194P}, \citealt{2009ApJ...703...96P},
\citealt{2010arXiv1010.2277R}, and \citealt{2010ApJ...713.1332R}). The
main idea is that magneto-thermal instabilities could lead to
different magnetic field configuration in the ICM depending on the
thermal properties of clusters.  When $\nabla T \cdot \vec{g}<0$,
i.e. when the gravity field $\vec{g}$ is anti-parallel to the
temperature gradient in the ICM, heat-driven flux buoyancy
instabilities (HBI) are developed. The HBI reorient the magnetic field
lines to be perpendicular to the temperature gradient, thus reducing
the effective conductivity of the plasma and preventing the cooling
catastrophe \citep{2010ApJ...713.1332R}. Recently,
\citet{2010ApJ...713.1332R} and \citet{2010ApJ...712L.194P} have shown
that modest levels of turbulence ($\sim $ 100 km/s) can suppress the
HBI, resulting in a quasi-stable thermal equilibrium, with
isotropically tangled magnetic field lines. However lower levels of
turbulence mixing are insufficient to suppress HBI, resulting in a
thermal runaway and leading to a cool core cluster. A different
magnetic field configuration would then be expected in clusters with
and without cool cores, and the combined effects of HBI and turbulence
would explain how minor and major mergers could disrupt the cool
core.\\ At the aim of quantify how much of the observed $F_P$ is due
to the gas density and magnetic field inside the core, we integrated
numerically Eq. \ref{eq:psiobs} from 0 to one core radius and from 1
to 10 core radii, assuming a magnetic field with $\langle B_0
\rangle=$5$\mu$G, $n_0=$0.017 cm$^{-3}$, $\beta=0.6$, $r_c=$90 kpc,
that are the weighted means for sources in the cool core sub-sample.
The ratio of the two quantities gives 0.79, indicating that the $RM$
resulting from the core is a consistent fraction of the total amount
which is responsible for the observed $F_P$.\\ The results we have
obtained in this section is not conclusive, but suggest a
possible difference in the magnetic field properties of clusters with
and without cool cores, thus supporting the scenario proposed by
\citet{2010ApJ...713.1332R} and
\citet{2010ApJ...712L.194P}. Unfortunately, these data are not
suitable for further investigation (i.e. to disentangle the effects of
the gas density and to better investigate if and how the difference in
the magnetic field properties of these two samples is due to different
magnetic field strengths and structure) but indicate that a more
detailed analysis of the magnetic field in these two samples could
offer important information about the interplay of magnetic fields,
thermal conduction and the bi-modality cool core, non cool core
clusters.\\%
\section{Magnetic field strength and cluster temperature}
\label{sec:BT}
Recent works have analyzed a possible connection between the magnetic
field strength and the cluster gas density and temperature. These
studies are based on cosmological simulations
(e.g. \citealt{2005xrrc.procE8.10D}, \citealt{1999A&A...348..351D},
\citealt{2005ApJ...631L..21B}) or plasma physical considerations
\citep{2010arXiv1003.2719K}.  SPH simulations predict that the mean
magnetic field strength varies with the ICM mean temperature according
to $B \propto T^2$. A shallower trend has been found by,
\citet{2010arXiv1003.2719K} who found $B \propto T^{3/4}$.\\ Here we
investigate a possible connection between the observed $F_P$ and the
cluster temperature. The selection criterion we have used, based on
high X-ray luminosity, naturally favors the selection of hot
clusters. Nonetheless, clusters in our sample span a good range of
temperatures, going from 3 to 13 keV.  We have divided our initial
sample in two sub-samples having $T >$ 7 keV and $T \leq$ 7 keV
respectively. The cut in temperature is set to have almost the same
number of sources in both samples.\\ The high and low T sub-samples
consist of 305 sources (163 upper limits) and 391 sources (171 upper
limits), respectively. The trend of $F_p$ versus $r_{norm}$ for these
two sub-samples is shown in Fig.  \ref{fig:DepoT_fede}. The logrank
test yields that the null hypotheses should be accepted with P$=$0.64,
thus indicating a common origin for the two population. The mean
central gas density $n_0$ for the two samples is 0.003$\pm$0.010
(T$>$7 keV sample) and 0.004$\pm$0.015 ($T \leq $ 7 keV sample),
meaning that the logrank test on $F_P$ is actually a test on the
magnetic field properties, while different values of the gas density
should not play an important role.\\ Recently
\citet{2010A&A...522A.105G} have analyzed the Faraday RM of a sample
of sources that belong to hot nearby galaxy clusters, and used these
data together with literature ones in order to investigate a possible
connection between the magnetic field strength and the cluster mean
temperature. Clusters are divided into three samples having T$<$4 keV,
4$\leq$T$\leq$8 keV and T$>$8 keV. The data analyzed do not show
evidence for such a correlation, indicating that a possible connection
between the magnetic field strength and the gas temperature, if
present, is very weak \citep{2010A&A...522A.105G}.  As previously
noted, the temperature threshold of 7 keV has been chosen arbitrarily
to divide clusters into two sub-samples.  In order to compare our
results with those of \citet{2010A&A...522A.105G}, we repeated the
test comparing clusters with T$\leq$4 and $T\geq$8 keV. There are 75
sources (28 upper limits) in clusters with T$\leq$4 keV and 305 sources
(163 upper limits) in clusters with T$\geq$8 keV. The logrank test
indicates that the null hypothesis can be accepted with
P$=$0.66. These data do not indicate a possible difference in the
magnetic field properties of clusters depending on their temperature,
but due to the small number of sources in the T$\leq$4 keV sample, a
more detailed analysis would be required. We note however that these
results are in agreement with those obtained by RM data (Govoni et
al. 2010).\\
\section{Conclusions}
\label{sec:concl}
We performed a statistical analysis of the fractional polarization of
radio sources in a sample of X-ray luminous galaxy clusters with the
aim of studying the properties of the intra-cluster medium magnetic
field. We used data from the NVSS to search for a trend in the
fractional polarization with distance from the cluster center,
following the approach proposed by \citet{2004A&A...424..429M}. The
sample of clusters we have selected comprises both cool core clusters
and merging systems. We have detected a clear trend of the fractional
polarization, being smaller for sources close to the cluster center
and increasing with increasing distance from the cluster centers. The
low fractional polarization in sources closer to the center is
interpreted as result of higher beam-depolarization, occurring because
of the higher magnetic field and gas density in these regions. This
result confirms that magnetic fields are ubiquitous in galaxy
clusters, as already found by \citet{2004JKAS...37..337C} and
\citet{2004mim..proc...13J}. Our results can be summarized as
follows:

\begin{itemize}
\item{We used 3D magnetic field simulations performed with the FARADAY
  code to search for the magnetic field model that best reproduces the
  trend of the fractional polarization with the distance from the
  cluster center. We assumed that the magnetic field power spectrum is
  Kolmogorov-like and that the magnetic field energy density decreases
  with radius as the gas thermal energy density. Under these
  assumptions, we found that a magnetic field with a central value of
  5$\mu$G gives the best agreement with the observed fractional
  polarization trend.}\\
\item{}We investigated possible differences between the $F_P$ trend
  observed in clusters with and without radio halos.  The logrank
  statistical test indicates that the two sample of sources very
  likely belong to the same intrinsic population (P=0.99). Magnetic
  fields in galaxy clusters are then likely to share the same
  properties regardless of the presence of radio emission from the
  ICM. This result poses problems for the ``hadronic-models'' for the
  origin of radio halos, that requires a difference in magnetic field
  strengths in clusters with and without radio halos.\\

\item{}We searched for possible differences in the magnetic field
  properties in clusters with and without cool core. The logrank test
  indicates that the $F_P$ distribution observed in clusters with and
  without cool cores is likely to be different (the null hypothesis of
  the two samples belonging to the same population has a low
  significance: P$=$0.15). This is expected by recently proposed
  models that explain the cool core and non cool core bimodality in as
  due to different magnetic field configurations in clusters with and
  without cool core (\citealt{2009ApJ...703...96P},
  \citealt{2010ApJ...712L.194P}, \citealt{2010arXiv1010.2277R}, and
  \citealt{2010ApJ...713.1332R}). The results obtained here must be
  taken with cautions since the role of the different gas densities in
  the two samples is not easy to quantify, and could play a crucial
  role. Deeper radio observations would be required to properly test
  these models.\\

\item{}We searched for a possible dependence of magnetic fields from
  the cluster mean temperature. Clusters were initially divided into
  two samples having T $\geq$ 7 keV and $<$ 7 keV. The result of the
  logrank test yields that the null hypothesis that the two samples
  are intrinsically drawn from the same population is accepted with
  P$=$0.64. In order to compare the $F_P$ analysis with the recent
  result by Govoni et al. (2010), we repeated the test by dividing the
  clusters into two samples having T$\geq$ 8 keV and T$\leq$4 keV. The
  logrank test indicates now that null hypothesis can be accepted with
  P$=$0.66. Due to the smaller number of sources in the T$\leq$4
  sample this result would need further investigation, but indicate
  that clusters with different temperatures share the same magnetic
  field properties.
%  is expected by theoretical works
%  (e.g. \citealt{2005xrrc.procE8.10D}, \citealt{1999A&A...348..351D},
%  \citealt{2005ApJ...631L..21B}, \citealt{2010arXiv1003.2719K}).
\end{itemize}
\bigskip
{\bf Acknowledgments} We thank L. Rudnick for very helpful discussions
and precious advices. We also thank V. Vacca, G. Brunetti and
U. Keshet for useful discussions, and the referee for his/her
  useful comments. The Synage++ package was used to perform part of
the analysis. This work was supported by the Italian Space Agency
(ASI), and by the Italian Ministry for University and research (MIUR).
A. B. and M. B. acknowledge support by the DFG Research Unit 1254
``Magnetization of interstellar and intergalactic media: the prospect
of low frequency radio observations''. This research has made use of
the NASA/IPAC Extragalactic Data Base (NED) which is operated by the
JPL, California institute of Technology, under contract with the
National Aeronautics and Space Administration.

\bibliographystyle{aa} \bibliography{master}

\appendix

\section{Investigating possible caveats}
\label{sec:app}
We will investigate here possible caveats that could affect our
analysis of the fractional polarization trend (Sec. \ref{sec:DP}). 
\subsection{Cluster sources and background sources}
We are analyzing the $F_p$ trend versus the cluster impact parameter,
so that in each bin we will have both sources that belong to the
cluster and sources that are in its background. The difference in the
$F_p$ experienced by a source lying in a plane perpendicular to our
line of sight and passing through the cluster center and a source that
is in the background of the cluster is a small factor (the dispersion
of the $RM$ changes of a factor $\sqrt{2}$), so that we are confident
that this should not cause a major effect. \\ Radio sources that
belong to the clusters in our sample could have larger angular sizes
with respect to distant background ones. In particular, cluster radio
sources sometimes show a ``narrow-angle tail'' or ``wide-angle tail''
morphology. Tails are known to be usually less polarized
intrinsically, i.e. when observed at high frequencies. In addition,
the ICM in the immediate surrounding of cluster radio sources could be
locally compressed, so that cluster radio sources could suffer higher
depolarization respect to background ones. Given that redshifts are
not known for all of the sources in our sample, a possible way to
distinguish background and cluster radio sources is to divide them
according to their angular size. We have separated the source sample
into two samples having size $>$5 beams and $<5$beams, and compared
the $F_P$ trends of the two populations. The value of 5 beams was
chosen on the basis of the radiosources in the cluster Abell 2255. In
this cases in fact the radio sources that are cluster members are
known (\citealt{2003AJ....125.2427M}), its redshift is representative
of the sample, and ``narrow-angle tail'' and ``wide-angle tail'' type
sources are present in the cluster. The logrank test indicates with
high significance (P$=$0.96) that the two samples have the same
properties, so that effects due to local compression of the ICM and
source size, as well as projection effects should not play any role.
We can then safely consider the sample as a whole, as we did.
\subsection{$F_P$ trend and cluster center.}
\begin{figure}[ht]
\centering
\includegraphics[width=0.9\columnwidth]{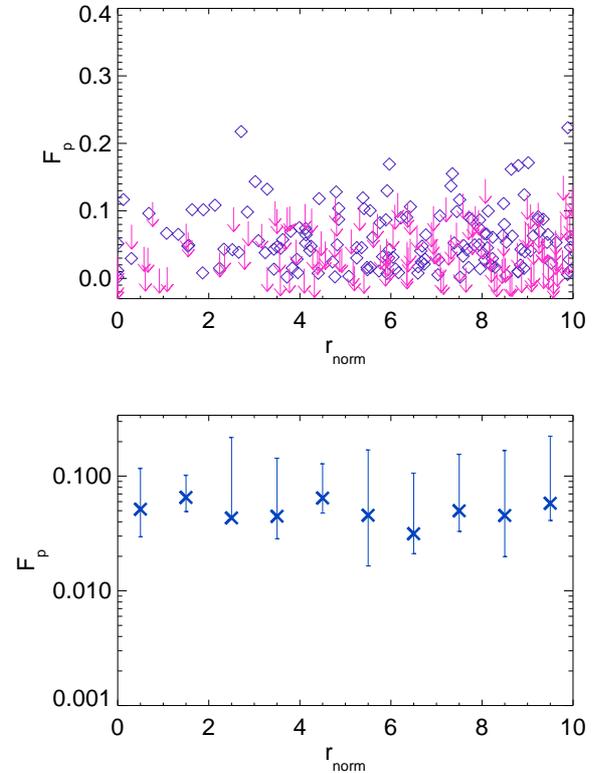}
\caption{ Fractional polarization of sources versus projected
  distance.  Sources are centered on a position that is 10$r_c$ far
  from the cluster center. Arrows indicate upper limits.  Bottom
  panel: crosses refer to the median of the KM estimator is each bin,
  bars indicate the 14th and 84th percentile. No trend is detected in
  this case }
\label{fig:testMatteo}
\end{figure}

Another issue is related to the binning process. The first bin,
$r_{norm}<1$ could show a lower $F_p$ because the number of sources
that fall within thin bin is  small, and the chance of finding
strongly polarized sources in then reduced. To verify this, we
repeated the same analysis described in Sec. \ref{sec:DP} but centered
on different position. The center is now chosen at a distance
corresponding to $\sim$10 $r_c$ for each cluster. The $F_p$
distribution is shown in Fig. \ref{fig:testMatteo}. No trend is
detected now. This guarantees that the trend detected in
Sec. \ref{sec:DP} is real and due to the cluster and not to missing
statistics.

\end{document}